\documentclass[usenatbib,useAMS]{mn2e}

\usepackage{amsfonts}
\usepackage{amsmath}
\usepackage{wasysym}
\usepackage{graphicx}
\usepackage{longtable}
 
\begin{document}

\label{firstpage}

\title[The IGIMF in ETGs]{The integrated galaxy-wide stellar initial mass function over the radial acceleration range of early-type galaxies}

\author[Dabringhausen \& Kroupa]{
J. Dabringhausen$^{1}$, \thanks{E-mail: joerg@sirrah.troja.mff.cuni.cz}
P. Kroupa$^{1,2}$ \\
$^{1}$ Charles University, Faculty of Mathematics and Physics, Astronomical Institute, V  Hole\v{s}ovi\v{c}k\'ach 2,\\
 CZ-180 00 Praha 8, Czech Republic\\
$^{2}$ Helmholtz-Institut f\"{u}r Strahlen- und Kernphysik, Universit\"at Bonn, Nussallee 14-16, 53115 Bonn, Germany}

\pagerange{\pageref{firstpage}--\pageref{lastpage}} \pubyear{2022}

\maketitle

\begin{abstract}
The observed radial accelerations of 462 Early-type galaxies (ETGs) at their half-mass radii are discussed. They are compared to the baryonic masses of the same galaxies, which are derived from theoretical expectations for their stellar populations and cover a range from $\approx 10^4 \, {\rm M}_{\odot}$ to $\approx 10^{11} \, {\rm M}_{\odot}$. Both quantities are plotted against each other, and it is tested whether they lie (within errors) along theoretical radial acceleration relations (RARs). We choose the Newtonian RAR and two Milgromian, or MONDian RARs. At low radial accelerations (corresponding to low masses), the Newtonian RAR fails without non-baryonic dark matter, but the two MONDian ones may work, provided moderate out-of-equilibrium dynamics in some of the low-mass ETGs. However all three RARs fail at high accelerations (corresponding to high masses) if all ETGs have formed their stellar populations with the canonical stellar initial mass function (IMF). A much better agreement with the observations can however be accomplished, if the theory of the integrated galaxy-wide stellar initial mass functions (IGIMFs) is used instead. This is because the IGIMF-theory predicts the formation of an overabundance of stellar remnants during the lifetime of the massive ETGs. Thus their baryonic masses today are higher than they would be if the ETGs had formed with a canonical IMF. Also the masses of the stellar-mass black holes should be rather high, which would mean that most of them probably formed when the massive ETGs were not as metal-enriched as they are today. The IGIMF-approach confirms downsizing.
\end{abstract}


\section[Introduction]{Introduction}
\label{sec:introduction}

Mass estimates for galaxies is an active field of research, where a number of surprising and intriguing discoveries have been made.

In spiral galaxies, rotation curves were measured to ever larger distances from the centres of these galaxies. It thereby turned out that the rotation curves flattened out \citep{Rubin1978,Bosma1981}. This stands in contrast to declining rotation curves at large distances, as the amount of directly observable matter predicts under the assumption of Newtonian dynamics. A straight-forward and popular explanation of this finding is that an additional component of unseen matter alters the potential of spiral galaxies, and thus their rotation curves \citep{Bosma1981,Rubin1985}. This includes the one of the Milky Way \citep{Bahcall1984a,Bahcall1984b}. 

Also the dynamics of elliptical galaxies can be interpreted as a consequence of unseen matter \citep{Cappellari2006,Bolton2008,Tortora2009}. The most extreme known cases are the large number of low-luminosity ETGs that surround the Milky Way, which exhibit optical mass-to-light ratios of up to some $10^3$ or even $10^4$ in Solar units, if they are assumed to be in virial equilibrium (e.g. \citealt{Mateo1998,Strigari2008,Wolf2010,Ackermann2014}).

For this reason, the low-luminosity ETGs are often interpreted as representing the most non-baryonic cold dark-matter (CDM) dominated galaxies in the Universe (e.g. \citealt{Strigari2008}). CDM in general would gather into haloes through gravitational collapse \citep{NFW1996,Moore1998,Gao2004}. Some CDM-haloes are thought to not even contain a galaxy (e.g. \citealt{Li2010}), while every primordial galaxy would be in a CDM-halo. However, the theory of General Relativity \citep{Einstein1916} is left unaltered with this approach, which leads eventually to the $\Lambda$CDM-model. The $\Lambda$CDM-model is summarized in \citet{Planckcoll2020}, but also see \citet{delPopolo2017}, \citet{Bullock2017} and \citet{Perivolaropoulos2021}.

Observations imply that low-mass ETGs are indeed very common inside and outside the Local Group \citep{Misgeld2008,Misgeld2009,Javanmardi2016}, while detecting the faintest ETGs, let alone measuring their dynamics, is still a serious observational challenge even in the Local Group. But in any case, the apparently rotationally supported disks of low-mass ETGs around the major galaxies in the Local Group cast serious doubts on the notion that the low-mass ETGs are primordial, CDM-dominated galaxies \citep{Kroupa2005,Metz2007,Metz2008,Pawlowski2012a,Ibata2013,Hammer2013,Pawlowski2013,Pawlowski2015,Pawlowski2020,Patel2020,
Pawlowski2021}. Disks of satellites are also a frequent feature beyond the Local Group \citep{Ibata2014,Ibata2015,Mueller2016,Mueller2018,Mueller2021}. The reason why they are problematic in the $\Lambda$CDM-model is that the $\Lambda$CDM-model predicts a predominately random distribution and motion of such primordial dwarfs around their host galaxies. As an alternative scenario, it has been suggested that low-luminosity ETGs formed out of the matter that was torn out of encountering galaxies through tidal forces, which makes the formation of rotating disks of satellites simply a consequence of the conservation of angular momentum \citep{Kroupa2005,Kroupa2010,Pawlowski2012b,Pawlowski2014,Pawlowski2018}. There is indeed strong evidence that many, if not most low-mass ETGs are tidal dwarf galaxies (e.g. \citealt{Okazaki2000,Kroupa2010,Dabringhausen2013}). Such tidal dwarf galaxies would however consist almost entirely of baryonic matter, even if their progenitor galaxies contained a substantial amount of CDM \citep{Barnes1992a,Duc2004,Bournaud2006,Bournaud2010}. Thus, the high internal velocity dispersions of low-luminosity ETGs are likely to have other reasons than a deepening of their potential wells through a presence of CDM.

An alternative approach to postulating the presence of CDM was made by \citet{Milgrom1983MOND}, who augmented the theory of gravitation in the limit of weak gravitational fields, instead of proposing new kinds of matter and energy. This modification is known as Modified Newtonian Dynamics (MOND), or Migromian Dynamics, and is one example of the modified gravity theories. MOND has passed many observational tests, as documented e.g. in the reviews by \citet{Famaey2012} and \citet{Banik2021}. Recent highlights are the possible detection of the external field effect (EFE) by \citet{Haghi2016,Chae2020,Chae2021} and assymetric tidal tails of open star clusters \citep{Kroupa2022}. The EFE is a phenomenon specific to modified gravity theories, but absent to the $\Lambda$CDM-model \citep{Chae2022}. 

The internal velocity dispersions of low-mass ETGs remain systematically too high also in MOND, when compared to their visible stellar mass, while not as much as in the $\Lambda$CDM-model \citep{Tortora2014,Dabringhausen2016b}. This could be a consequence of out-of-equilibrum dynamics in Newtionian dynamics \citep{Kroupa1997,Casas2012}, as well as in Milgromian dynamics \citep{McGaugh2010}. 

A missing mass problem is also extensively documented for the ETGs at the bright end of the galaxy luminosity function, even though the mass discrepancy detected for them is not as spectacular as the one in spiral galaxies or low luminosity ETGs. However, the question is whether the ETGs all have a universal stellar initial mass function (IMF) and a non-baryonic dark matter component comparable to the mass in stars, or whether the IMF itself changes from ETG to ETG.

The reference model for a non-changing, canonical IMF is given as
\begin{equation}
\xi (m) =k \, k_i \, m^{-\alpha_{i}},
\label{eq:IMF}
\end{equation} 
with 
\begin{align}
\nonumber \alpha_{1} = 1.3  & {\rm \ if \ } 0.1 \le  \frac{m}{\rm{M}_{\odot}} < 0.5,\\
\nonumber \alpha_{2} = 2.3  & {\rm \ if \ } 0.5 \le  \frac{m}{\rm{M}_{\odot}} < 1.0,\\
\nonumber \alpha_{3} = 2.3  & {\rm \ if \ } 1.0 \le  \frac{m}{\rm{M}_{\odot}} \le m_{\rm max}
\end{align}
\citep{Kroupa2001,Kroupa2013}, or a function that cannot be distingushed from it with observations (e.g. \citealt{Chabrier2003,Dabringhausen2008}). In the above function, $m$ is the initial stellar mass, $m_{\rm max}$ is the maximum initial stellar mass, the factors $k_i$ ensure that the IMF is continuous where the power changes and $k$ is a normalization constant. The parameter $k$ ensures that the integral over the IMF equals unity, even if the other parameters change. There is some debate over the correct value for $m_{\rm max}$ in equation~\ref{eq:IMF}, as is evident by the differences in $m_{\rm max}$ in the papers by, for example, \citet{Bruzual2003} ($m_{\rm max}=100 \, {\rm M}_{\odot}$), \citet{Weidner2004ul} ($m_{\rm max}\approx 150 \, {\rm M}_{\odot}$) and \citet{Crowther2010} ($m_{\rm max}\approx 300 \, {\rm M}_{\odot}$). This has however almost no consequences for mass estimates from equation~\ref{eq:IMF}, since for this value of $\alpha_3$, the total number of stars with masses $m>100 \, {\rm M}_{\odot}$ is extremely small compared to stars with lower masses.

The reason why such a stellar population is so popular as a reference is that, for a long time, it appeared that equation~\ref{eq:IMF} is consistent with observations of all stellar populations in the Milky Way \citep{Kroupa2001,Kroupa2013}. It is thus remarkable that the virial masses of high-luminosity ETGs can in general not be explained with it. This is shown e.g. in \citet{Cappellari2006,Bolton2008,Tortora2009,Cappellari2012,Samurovic2014} and \citet{Dabringhausen2016b}, even though these authors discuss different solutions to the problem (i.e. changing IMFs, CDM or MOND, and sometimes more than one of these).

If no non-baryonic dark matter is to be invoked, the solution may be an IMF that changes from galaxy to galaxy, which leads to the theory of the integrated galaxy-wide stellar initial mass functions (IGIMFs). The IGIMF-theory is based on the conjecture that all stars form in groups or embedded star clusters, and never in isolation \citep{Kroupa1995b}. All the stars that form in the different star clusters of a galaxy over a certain time make up its IGIMF \citep{Kroupa2003}, and the mass of the most massive star in a specific star cluster depends on the mass of the star cluster \citep{Weidner2010,Yan2017}. The mass of the most massive star cluster in a galaxy depends in turn on the star formation rate (SFR) of the galaxy \citep{Weidner2004sfr}, and already from this, it is evident that no universal galaxy-wide IMF, or IGIMF can exist among galaxies. However, \citet{Dabringhausen2008,Dabringhausen2009,Dabringhausen2012} and \citet{Marks2012imf} found that also the high-mass slope of the IMF in star clusters, i.e. $\alpha_3$ in equation~\ref{eq:IMF}, depends on their mass, so that it is shallower in more massive star clusters. Thus, very massive star clusters contain more massive stars not only in total, but also per unit of low-mass stars. Moreover, \citet{Marks2012imf} found an increase of low-mass stars with the metallicity of the star clusters. After some more studies on the variations of stellar populations among galaxies due to the IGIMF-theory \citep{Weidner2011,Weidner2013a,Weidner2013b,Fontanot2017,Yan2017}, this was put together by \citet{Jerabkova2018} into a grid of modelled stellar populations, whose high-mass IMF slopes and most massive stars depends on their SFR, and whose low-mass IMF depends on their metallicity. These changes in the stellar populations can explain the overabundance of low-mass stars in massive, and thus metal-rich ETGs (e.g. \citealt{vanDokkum2010,Cappellari2012,LaBarbera2013,LaBarbera2019,Smith2020,Gu2022}). Other studies on the same topic revealed that there is a gradient in the IMF of the massive ETGs in such a way, that it was extremely over-abundant in low-mass stars especially at their centres, while in their outer parts it approaches the canonical IMF (e.g. \citealt{LaBarbera2016,vanDokkum2017,Parikh2018,Sarzi2018,Bernardi2018,DominguezSanchez2019,LaBarbera2019}). In the context of the IGIMF-theory, this could mean that present-day star formation with high metallicites happens mainly at the centre of the ETGs, whereas earlier on the star forming regions were more extended. Also a lack of high-mass stars was detected in low-mass late-type galaxies \citep{Lee2009}. The IGIMF-theory can also explain an overabundance of massive stars in the most massive star clusters (UCDs, \citealt{Dabringhausen2008,Dabringhausen2009,Dabringhausen2012,Marks2012imf}). Finally, it can explain the colour dependency of star-forming disk galaxies \citep{Lee2009,Gunawardhana2011}, or their element abundances and their ratios \citep{Romano2017,Zhang2018,Yan2020,Yan2021}. With a universal IMF for all galaxies, all of this would be difficult, if not impossible.

In other words, we propose the IGIMF-model as the common model for how stellar populations emerge from star formation, in dependence of the size, star-formation rate and metallicity of a given star-forming gas-cloud. This has notably consequences for the massive ETGs, for which many of the above authors claim that their IMFs in their centres are exclusively bottom-heavy. According to the IGIMF-model, this is only true for their most recent star-formation events, which have moderate SFRs and are already enriched with metals. In their past however, when they were less enriched with metals and their star formation rates were high, their IGIMFs were top-heavy according to the IGIMF-model. Massive stars burn out fast, and contribute then only the mass of their remnants to the total masses of their ETGs. Thus, in total, massive ETGs are heavier than what can be explained with a purely canonical IMF because of a combination of bottom-heavy star formation today and top-heavy star-formation in the past. This could in principle be checked by testing whether the line indices of the low-mass stars propose enough low-mass stars to account for the missing mass in total \citep{Kroupa1994}. That additional missing mass might be needed, despite the proposed bottom-heaviness of the IMF, was already hinted at in \citet{vanDokkum2010}.

In this contribution, we will deal with the mass estimates for ETGs with and without the IGIMF-theory. Most of the massive ETGs have formed their stellar populations Gyrs ago, which implies that massive stars have already evolved into remnants in them. In consequence, these stars no longer contribute significantly to the optical luminosity of the ETGs, while the remnants still contribute to their mass. Efforts to link the star formation histories of ETGs with their present-day masses have shown that the characteristic timescale for building up their stellar populations decreases with increasing mass \citep{Thomas2005,Recchi2009,Yan2021}. Thus, the most massive (and most luminous) ETGs also had the highest SFRs in the past. Therefore the most luminous ETGs should, according to the IGIMF model, have the most stellar remnants per unit mass of low-mass stars. This could in principle explain the mass deficit detected for them, if their IGIMF is wrongly assumed to be identical to the IMF given in equation~(\ref{eq:IMF}). We will however not deal with the metallicity dependence of the IGIMF, as the luminosity of the ETGs is an observable. It came to be in its present magnitude whatever the history of a galaxy was, including changes of its IGIMF with metallicity. In this sense, the luminosity is similar to the internal velocity dispersions and the half-light radii, which are the observational basis for the dynamical masses of the galaxies, i.e. the other key observational parameter of the galaxies in this paper. 

Thus, the question is probed whether the mass-discrepancy evident in ETGs disappears, if the independently formulated IGIMF model is adopted to quantify the stellar populations in ETGs. Newtonian gravitation and MOND with either the standard $\mu$-function \citep{Milgrom1983gal} or the simple $\mu$-function \citep{Famaey2005} are tested in this paper. For this, the concept of radial acceleration relations (RAR) is used, which are discussed by e.g. \citet{Wu2015,McGaugh2016} or \citet{Lelli2017}. RARs emerge essentially by the comparison between the acceleration implied by the observed dynamics of galaxies with the acceleration expected based on the matter density profile of the same galaxies. The latter is at least to some extent based on models and assumptions (such as virial equilibrium, the mass-to-light ratio of the stellar population, the presence of an CDM-halo, etc), and the RAR thereby becomes a powerful tool to test such models and assumptions.

This paper is organized as follows. Section~(\ref{sec:data}) describes the specific data on ETGs from \citet{Dabringhausen2016a} that are used for linking IGIMF-related parameters of ETGs to their observed properties. Section~(\ref{sec:methods}) introduces the IGIMF-model, including some adjustments and parameterisations of the IGIMF introduced in \citet{Dabringhausen2019}, and used in this paper. In Section~(\ref{sec:results}) our results are laid out, and a discussion is given in Section~(\ref{sec:discussion}). This is followed by a summary and conclusion in Section~(\ref{sec:conclusion}).

\section{Data}
\label{sec:data}

\subsection{Selection criteria for the sample of ETGs}
\label{sec:selection}

The data that are used to link the IGIMF to observed early-type galaxies are provided in the catalogue by \citet{Dabringhausen2016a}. This catalogue comprises 1715 ETGs, which span the whole luminosity range of ETGs from faint dwarf spheroidal galaxies (mostly from the catalogue by \citealt{McConnachie2012}) to giant elliptical galaxies (to a large extent from the ATLAS$^{\rm 3D}$ survey, \citealt{Cappellari2011}). What motivates to combine these galaxies into a single catalogue, and similar galaxies from other sources as well, is that they share two properties: There is little, if any, star formation in them at present, and random motion dominates over ordered motion for their stellar populations. Apart from these two defining properties, the properties of the ETGs are diverse. However, ETGs gather close to a (one-dimensional) line in the (two-dimensional) mass-radius space. This is similar to main-sequence stars (see e.g. \citealt{Demircan1991} for the stars).

The quantities from that catalogue that are relevant for the present paper are the effective half-light radii of the ETGs, $R_{\rm e}$, their S\'{e}rsic indices, $n$, their observed central line-of-sight (LOS) velocity dispersions, $\sigma_0$, the rotational velocities within their effective radii, $v_{\rm rot}$, the masses of their stellar plus remnant populations under the assumption that the IMF is canonical, $M_{\rm can}$, the luminosity in the $V$-band, $L_V$. Also a time, $t$, that represents the overall age of the stellar population and a metallicity, $Z$, that gives the average metal content of the stars are collected from \citet{Dabringhausen2016a}. These ages and metallicities are provided by the literature that \citet{Dabringhausen2016a} used, and are thus subject to the assumptions made there. The parameters $R_{\rm e}$, $n$ and $\sigma_0$ are needed for estimating the dynamical mass of the galaxies, which are in turn needed for estimating a characteristic radial acceleration that particles seem to experience based on their dynamics, $a_{\rm dyn}$. $M_{\rm can}$ is not only needed for comparing it to $M_{\rm dyn}$, but also for estimating their SFR at the time when the majority of their stars formed, $SFR_{\rm peak}$. $L_V$ plays a crucial role in the estimate of $M_{\rm can}$. The availability of these data restrict the catalogue from 1715 ETGs to 462 ETGs, which however still cover the almost whole luminosity range of ETGs, from $L_V \approx 10^4 {\rm L}_{\odot}$ to $L_V \approx 10^{11} {\rm L}_{\odot}$.

The data on $v_{\rm rot}$ are also collected, since rotation may provide a notable contribution to the overall kinetic energy of the matter in the ETGs. Thus, it is relevant for mass estimates based on the virial theorem. If also $v_{\rm rot}$ is considered, the number of galaxies discussed is limited from 462 to 310. However, the availability of $v_{\rm rot}$ is not made a requirement for a galaxy to be discussed in this paper. We rather examine the full sample at the expense of ignoring $v_{\rm rot}$, and a subsample, where $v_{\rm rot}$ is considered. This is also done because the galaxies, for which $v_{\rm rot}$ was not measured, lie mostly at low luminosities and masses. Note that $v_{\rm rot}$ in the subsample may still be consistent with zero; the important point is that they were measured.

We refer the reader to \citet{Dabringhausen2016a} for a detailed description of the full catalogue. Additional Information on the restrictions made on this catalogue, so that only 462 galaxies remain for the present paper, are given in \citet{Dabringhausen2016b} and \citet{Dabringhausen2019}. While the subset of 462 galaxies used in this paper is exactly the same as in \citet{Dabringhausen2019}, some specifics are especially relevant here. They will be detailed in this Section.

\subsection{Luminosities}
\label{sec:data-luminosities}

\citet{Dabringhausen2016a} collected luminosities in many passbands, and \citet{Dabringhausen2019} compared these data from different passbands to each other. Here on the other hand, we restrict ourselves to the data on the $L_V$ passband. The reason is that the sample on $L_V$ is the largest data sample among the luminosities in \citet{Dabringhausen2016a}, from direct measurements of individual galaxies as well as from derivations from statistical data. Choosing e.g. the $I$ passband instead would narrow the remaining ETGs from 462 to 347.
 
\subsection{Stellar masses and metallicities of the stellar populations} 
\label{sec:data-masses}

The estimates on $M_{\rm can}$ given in \citet{Dabringhausen2016a} are based on a large set of models for simple stellar populations (SSPs) by \citet{Bruzual2003}. These SSPs are defined as stellar populations that have formed instantly at a certain time with a certain metallicity. 

\citet{Dabringhausen2016a} obtained $M_{\rm can}$ of a given ETG by first searching an SSP-model by \citet{Bruzual2003} that represents the age and the colour of the ETG the best. For the ETGs considered in this paper, estimates for age and metallicity were already given in the source papers of \citet{Dabringhausen2016a}. Thus, they simply adopted the $M_{\rm can}/L_V$ predicted by the according SSP-model as the $M_{\rm can}/L_V$ of the ETG. Finally, they multiply this $M_{\rm can}/L_V$ by the $L_V$ of the ETG.

There are certainly more elaborate methods to determine the stellar mass $M_{\rm can}$ of a ETG. For instance, the less massive the ETG is, the longer star formation tends to last in it (e.g. \citealt{Thomas2005} and \citealt{Weisz2014}). Thus, instant star formation is becoming an ever rougher sketch to their true star formation history the less massive the ETG is. In such cases, the adopted ages and metallicities can imagined best as characteristic values. On the other hand, this paper does resolve the age-metallicity degeneracy \citep{Worthey1994} to some extent, as it leaves the ETGs some freedom regarding their best-fitting ages and metallicities. This stands in contrast to some other rather recent papers, which keep the age fixed (e.g \citealt{Forbes2008,Forbes2011,Misgeld2011}), or are even equivalent to keeping age and metallicity fixed (e.g. \citealt{McConnachie2012} by assuming a mass-to-light ratio of one in Solar units for every galaxy in his sample).

Recall however the main purpose of this Paper. It is {\it not} to estimate the masses of individual galaxies as precisely as possible. For this, the method of fits by multiple SSPs by e.g. \citet{Blanton2007} or \citet{Cappellari2013b} would evidently be better than the fit to only a single SSP done here, but also far more complicated. Instead, our goal is to show that with MOND and the IGIMF, it is much better possible to understand the dynamics of the {\it ensemble} of ETGs without non-baryonic dark matter, in contrast to without these concepts. 

\subsection{Effective half-light radii}
\label{sec:data-radii}

The projected half-light radii, $R_{\rm e}$, published in \citet{Dabringhausen2016a} are based on observed values from a multitude of papers. If angular radii are published in their sources, \citet{Dabringhausen2016a} transform them into pc using the distance estimates they adopted for the galaxies in their catalogue. Values for $R_{\rm e}$ given in parsec in the source papers are transformed by \citet{Dabringhausen2016a} into angular radii using the distance estimates given in the source papers, and then transformed back into parsec again using the respective distance estimates listed in \citet{Dabringhausen2016a}.

Most $R_{\rm e}$ for the massive galaxies come initially from the ATLAS$^{\rm 3D}$-study \citep{Cappellari2011}. More specifically, \citet{Dabringhausen2016a} took the $R_{\rm e}$ from \citet{Cappellari2013}, which is the same source they used for the velocity dispersions of the respective ETGs. However, \citet{Cappellari2013} only took the luminosities they could observe within their fields of view, and calculated their $R_{\rm e}$ on this basis. As a matter of principle, these values for $R_{\rm e}$ are somewhat smaller than values for $R_{\rm e}$, which are calculated from the luminosities of the ETGs with infinite radii.
	
Finding out the correct $R_{\rm e}$ for massive ETGs is a tricky business. When \citet{Cappellari2013} compared their values for $R_{\rm e}$ with the values they have obtained for the $R_{\rm e}$ of the same ETGs in \citet{Cappellari2011}, they find that they have to multiply the new values by a factor of 1.35 in order to match the old values. However, the fit is excellent then. The reason is probably that the $R_{\rm e}$ in \citet{Cappellari2011} are normalised such that they correspond on average to the reference catalogue by \citet{deVaucouleurs1991}, who extend the radii of their galaxies to infinity. \citet{Cappellari2013} on the other hand want to avoid the assumptions involved in extending the luminosity profile to infinity, and cut the ETGs to the field of view of the telescope they used. We nevertheless multiply the $R_{\rm e}$ from \citet{Cappellari2013} with 1.35, in concordance to what they did for their analysis as well. 

Other contributions to the $R_{\rm e}$-values of massive ETGs in \citet{Dabringhausen2016a} come from \citet{Bender1992} and \citet{Scodeggio1998}, even though minor compared to the contribution by \citet{Cappellari2013}. In fact, the contribution by \citet{Bender1992} disappears, if also information on rotations is required (see Section~\ref{sec:data-rotation}). However, in contrast to the data on $R_{\rm e}$ from \citet{Cappellari2013}, the $R_{\rm e}$ from these sources are left unchanged, because the luminosity profiles were already extrapolated to infinity in the source papers. 

Also for the remaining ETGs, which are almost exclusively dwarf galaxies, we take the values for $R_{\rm e}$ unchanged, for reasons explained in Section~\ref{sec:data-sersic}.

\subsection{S\'{e}rsic indices}
\label{sec:data-sersic}

The Sersic index \citep{Sersic1968} is the number $n$ that best describes the overall luminosity profile of the galaxy in a generalised luminosity profile of the form
\begin{equation}
I(R)=\frac{L}{R_{\rm e}^2} \frac{b^{2n}}{2 \pi n \Gamma (2n)} \exp \left[-b_{n} \left(\frac{R}{R_{\rm e}}\right)^\frac{1}{n}\right],
\label{eq:sersic-profile}
\end{equation}
where $I(R)$ is the surface brightness of the galaxy as a function of the projected radius $R$, $I_0$ is the central surface brightness, $R_{\rm e}$ is the effective radius and $\Gamma$ is the gamma-function (e.g. \citealt{Sersic1963,Bernardi2018}). The parameter $b_{n}$ can be approximated as $b_{n} \approx 2n-0.324$ \citep{Ciotti1991}.

In the catalogue by \citet{Dabringhausen2016a}, $n$ is adopted from the literature they used, if available there. If not, they calculate $n$ from $R_{\rm e}$, using
\begin{equation}
\log_{10} (n) = 0.28+0.52 \log_{10}\left(\frac{R_{\rm e}}{{\rm kpc}}\right),
\label{eq:sersic}
\end{equation}
which is a relation that \citet{Caon1993} derived from observational data.

Thus, $n$ grows with the $R_{\rm e}$ of an ETG, which in turn tends to get larger with its mass. So especially for the massive ETGs with their larger $R_{\rm e}$ and $n$, the density profiles of the ETGs stretch more to the outer regions. This is even the case if $R_{\rm e}$ would remain the same with growing $n$, as is apparent from figure~1 in \citet{Ciotti1991}. Because of this behaviour of $n$, it is particularly the large ETGs with $n \apprge 4$ that must be covered fully by their observations, or else their $R_{\rm e}$ would appear smaller than they actually are. \citet{Cappellari2013} have indeed found a correction factor by which they have to multiply their massive ETGs, which they have observed only to finite radii. Otherwise they would not match on average the values suggested for the same galaxies with unlimited radii (\citealt{deVaucouleurs1991,Cappellari2011}, Section~\ref{sec:data-radii} in this paper).

For the other ETGs, their luminosity profiles have either been extrapolated to infinity in their sources, or they are dwarf galaxies. Dwarf ETGs mostly have $1 \apprle n \apprle 2$, and are thus more centrally concentrated than the massive ETGs (see figure~1 in \citealt{Ciotti1991}). Therefore, cutting off the outer parts of the dwarf galaxies is much less problematic than with the massive ETGs, and for this reason, also their $R_{\rm e}$ are adopted unchanged from the catalogue by \citet{Dabringhausen2016a}.

\subsection{Velocity dispersions}
\label{sec:data-velocities}

In the context of the present paper, we have to distinguish carefully between the expected line-of-sight (LOS) velocity dispersion of the tracer population due to the gravitational potential of the ETG, and the observed LOS velocity dispersion. We denote the former as $\sigma_0$, and the latter as $\sigma_{\rm obs}$. $\sigma_0$ is completely unordered motion, while $\sigma_{\rm obs}$ is not always. Also the two quantities are often used interchangeably, while they not always are.

As an example for $\sigma_0 \neq \sigma_{\rm obs}$, consider a tracer population of binary stars. In this case, $\sigma_{\rm obs}>\sigma_0$, because the stars in the binary move around their common centre of mass, while the binary systems as wholes move due to the potential of the galaxy. The former is completely unrelated to the potential of the galaxy, but nevertheless adds to $\sigma_{\rm obs}$.

Now consider a rotating galaxy, or more generally, a galaxy where the mean velocity towards the observer changes over the field of view. Also this case implies $\sigma_{\rm obs}>\sigma_0$, because $\sigma_0$ only quantifies the unordered motion, while $\sigma_{\rm obs}$ stems from both the ordered and the unordered motion within the field of view. The effect from ordered motion can in theory be eliminated by decreasing the field of view until only the centre of the ETG is observed, but this is in practice only possible up to a certain point. However, in contrast to binaries, rotation has the same cause as the unordered motion, namely the overall gravitational field of the galaxy. Therefore it cannot be separated easily from $\sigma_0$.

\citet{Dabringhausen2016a} follow the usual convention by designating the central LOS velocity dispersions of the ETGs as $\sigma_0$, while the quantities that they actually list in their catalogue are values for $\sigma_{\rm obs}$ according to the above definitions for $\sigma_0$ and $\sigma_{\rm obs}$. Their data on $\sigma_{\rm obs}$ are taken from the literature they use, if such a value was published there. If only the {\it average} observed LOS velocity dispersion within $R_{\rm e}$, $\sigma_{\rm e}$, is available from the literature they use, they estimate $\sigma_{\rm obs}$ using the relation
\begin{equation}
\log_{10}\left(\frac{\sigma_{\rm obs}}{{\rm km/s}}\right)=1.0478 \, \log_{10}\left(\frac{\sigma_{\rm e}}{{\rm km/s}}\right)-0.0909,
\label{eq:sigmae}
\end{equation}
which is obtained from a least-squares fit to data on 260 ETGs in \citet{Cappellari2013}, see the upper panel of their figure~8. \citet{Dabringhausen2016a} thus take the data on the velocity dispersions within the central parsec of the ETGs in \citet{Cappellari2013} as a measure for $\sigma_{\rm obs}$. Equation~(\ref{eq:sigmae}) can therefore not be used on ETGs with $R_{\rm e}<1 \, $kpc by construction, which is why \citet{Dabringhausen2016a} set $\sigma_{\rm obs}=\sigma_{\rm e}$ for such ETGs. This is well motivated by the findings that $R_{\rm e}=1$ kpc corresponds to $n \approx 1.9$ according to equation~(\ref{eq:sersic}), and that the velocity dispersion profiles of galaxies with $n\apprle 2$ are almost flat within their $R_{\rm e}$ (see \citealt{Graham1997}, their figure~8).

How relevant rotation (i.e. ordered motion) and binaries actually are for mass estimates of ETGs will be discussed in more detail in Sections~(\ref{sec:bin}) and~(\ref{sec:rot}).

\subsection{Rotational velocities}
\label{sec:data-rotation}

\citet{Dabringhausen2016a} aim at giving the rotational velocity of the ETGs, $v_{\rm rot}$, as the average rotational velocity of the ETG within $R_{\rm e}$, but the provided data in the source literature is quite heterogeneous. Actual estimates for the average rotational velocity are mainly given for luminous ETGs, whose extension and surface brightness allow to take independent spectra from many patches within $R_{\rm e}$, and derive the underlying motions for each of these spectra. Less luminous galaxies are at a given distance more challenging to observe in detail than their more luminous counterparts. Estimates of their rotation therefore tend to be based on fewer individual telescope pointings, and are usually rather quantified through maximum rotational velocities or the rotational velocities at $R_{\rm e}$ instead of proper averages within $R_{\rm e}$.

The data for $v_{\rm rot}$ provided by \citet{Dabringhausen2016a} may however be more comparable than it may seem from the inhomogeneities in the data in the source papers. By construction, the average rotational velocity is smaller than the maximum rotational velocity, which however is reached well within $R_{\rm e}$ for many low-luminosity ETGs (\citet{Geha2003}). Therefore, the average rotational velocity may for these ETGs actually be quite close to the maximum rotational velocity. On the other hand, there are also low-luminosity ETGs for which the rotational speed may still increase outside $R_{\rm e}$. Thus, an estimate for the rotational velocity for such ETGs at the radius $R_{\rm e}$ is smaller that the maximum rotational velocity, but is again probably quite close to it.

\section[Methods]{Methods}
\label{sec:methods}

\subsection{The IGIMF...}
\label{sec:IGIMF}

The fundamental assertions for the IGIMF is that most, if not all stars form in groups and not in isolation \citep{Kroupa1995b}, and that the IMFs in these groups are not universal, but depend on star formation conditions.

The metallicity determines the abundance of low-mass stars in IMFs in such a way that their number increases with metallicity. \citet{Kroupa2002} and \citet{Marks2012imf} use the equation
\begin{equation}
\alpha_i=\alpha_{ic}+\Delta \alpha \times[{\rm Fe}/{\rm H}]
\end{equation}
for $i=1$ and $i=2$ in equation~\ref{eq:IMF} for this, i.e. for stellar masses $m \le 1 {\rm M}_{\odot}$. $\alpha_{ic}$ are the low-mass IMF slopes of the canonical IMF (equation~\ref{eq:IMF}). For $\alpha_3$ in equation~\ref{eq:IMF}, i.e. for $m > 1 {\rm M}_{\odot}$, \citet{Marks2012imf} find
\begin{equation}
\label{eq:alpha3}
\alpha_3=
\begin{cases}
2.3 & {\ \rm if \ } x \ < -0.87\\
1.94-0.41 \, x & {\ \rm if \ } x \ \ge -0.87,
\end{cases}
\end{equation}
with
\begin{equation}
x=-0.14 \times [{\rm Fe}/{\rm H}]+0.99 \times \log_{10} \left( \frac{\rho_{\rm cl}}{10^6 \, {\rm M}_{\odot} \, {\rm pc}^{-3}}\right).
\end{equation}
In the above equation, $\rho_{\rm cl}$ is the density of the embedded star cluster, i.e. of the star-forming molecular cloud core, or in other words, the infant star cluster still containing the left-over gas besides the stars. Thus, the number of massive stars mostly increases with $\rho_{\rm cl}$, corrected by a small decrease with rising metallicity. The equation
\begin{equation}
\label{eq:rh-mass}
\frac{r_{\rm h}}{{\rm pc}}=0.10 \times \left( \frac{M_{\rm ecl}}{{\rm M}_{\odot}} \right)^{0.13},
\end{equation}
connects $\rho_{\rm cl}$ with the mass of the embedded cluster, $M_{\rm ecl}$, over its average 3D half-light radius, $r_{\rm h}$. The mass of the most massive star cluster that can form in a galaxy, $M_{\rm ecl,max}$, depends on the SFR of the galaxy as
\begin{equation}
\label{eq:SFR-clustermass}
 \frac{M_{\rm ecl,max}}{{\rm M}_{\odot}}=84793\times\left(\frac{{\rm SFR}}{{\rm M}_{\odot} \, yr^{-1}}\right)^{0.75},
\end{equation}
according to \citet{Weidner2004sfr}. The star cluster mass determines in turn the mass of its most massive star \citep{Weidner2010,Yan2017}. A parametrisation of the dependence of the mass of the most massive star, $m_{\rm max}$, on $M_{\rm ecl}$ is
\begin{align}
\log_{10}(m_{\rm max})  & = 2.56 \log_{10}(M_{\rm ecl}) \nonumber \\ 
& \times (3.82^{9.17} + [\log_{10} (M_{\rm ecl})]^{9.17})^{(1/9.17)}-0.38,
\label{eq:Mmaxfit}
\end{align}
\citep{Pflamm2007}. All star clusters and stellar groups of a galaxy with their different IMFs produce over a certain time its IGIMF, as introduced by \citet{Kroupa2003}. The shortest time span that is thought to populate the IGIMF of a galaxy completely is about 10 Myr \citep{Weidner2004sfr}, the average lifetime of molecular clouds.

\subsection{...and its simplification}
\label{sec:simplify}

\citet{Jerabkova2018} collected this information into a grid of IMFs that depend on the metallicities in which the stars form and on the SFR, and plotted a sample of these in their figure~1. Generally, the metallicity and the star formation in ETGs are low at first, then both rise, and finally the SFR is low again while the metallicity is high. This is shown in fig.~2 in \citet{Jerabkova2018},  which illustates that indeed a few star clusters form with IMFs like in the snapshots in figure~1 with extreme metallicities early in the life of the galaxy, but many more with IMFs more akin the canonical IMF later on.

More importantly is however that the quantities that are looked at in this paper are rather insensitive to the exact shape of the resulting IGIMF. This is notably also the case at its lower end, where the IGIMF is especially sensitive to the metallicity. We are however not interested in colours (like e.g. \citealt{Gunawardhana2011}) or ratios of line indices (like e.g. \citealt{vanDokkum2010}) in order to measure the shape of the low-mass IGIMF. Instead, the $L_V$-luminosities are the only quantity connected to the light of the galaxies that we care about, independently of the slope or shape of low-mass IGIMF that produces it. It is therefore safe in the context of this paper to set the IGIMF to constant values below masses of 1 ${\rm M}_{\odot}$ for a simplification.
 
Above masses of 1 ${\rm M}_{\odot}$, there is a weak dependency on metallicity as well, but we neglect it over the much stronger dependency on the SFR in this mass range. Thus, as in \citet{Fontanot2017}, we consider for this paper the IGIMF on its whole mass-range only as a function of SFR of a galaxy, and not on its metallicity. 

\citet{Dabringhausen2019} simplified the IGIMFs of the ETGs even further to
\begin{equation}
\xi_{\rm IGIMF} (m) =k \, k_i \, m^{-\alpha_i},
\label{eq:IGIMF3}
\end{equation}
with 
\begin{align}
\nonumber \alpha_1 = 1.3  & {\rm \ if \ } 0.1 \le  \frac{m}{\rm{M}_{\odot}} < 0.5,\\
\nonumber \alpha_2 = 2.3  & {\rm \ if \ } 0.5 \le  \frac{m}{\rm{M}_{\odot}} < 1.0,\\
\nonumber \alpha_{\rm eff} \, \in \, \mathbb{R}  & {\ \rm if \ } 1.0 \le  \frac{m}{\rm{M}_{\odot}} \le m_{\rm max},
\end{align}
where $m$ is the initial stellar mass, $m_{\rm max}$ and $\alpha_{\rm eff}$ are SFR-dependent parameters, the factors $k_i$ ensure that the IGIMF is continuous where the power changes and $k$ is a normalization constant that ensures that the integral over the IGIMF equals unity, even if the other parameters change. Although this IGIMF looks from its parametrisation very similar to the IMF given in equation~\ref{eq:IMF}, it is based on a completely different concept.

The values by \citet{Fontanot2017} for the SFR-dependent mass of the most massive stars, $m_{\rm max}$, is parametrized by \citet{Dabringhausen2019} as
\begin{align}
\label{eq:interpolationMmax}
\nonumber \frac{m_{\rm max}}{{\rm M}_{\odot}} = & \left[\log_{10}\left(\frac{SFR}{{\rm M}_{\odot}{\rm yr}^{-1}}\right)+5.824\right]^{3.133}\\
- & 0.4086\times \log_{10} \left(\frac{SFR}{{\rm M}_{\odot}{\rm yr}^{-1}}\right)
\end{align}
for $\log_{10}(SFR/{\rm M}_{\odot}{\rm yr}^{-1})<-0.878$, and $m_{\rm max}=150 \ {\rm M}_{\odot}$ for $SFR \ge -0.878 \ {\rm M}_{\odot} \, {\rm yr}^{-1}$. Thus, the physical mass limit for stars is set here to 150 ${\rm M}_{\odot}$, in concordance with \citet{Weidner2004ul} and \citet{Oey2005}. Even more massive stars may form \citep{Crowther2010} from the mergers of binaries \citep{Banerjee2012a,Banerjee2012b}. Their frequency is strongly suppressed, though, as also their progenitor stars are rare by the power $\alpha_3$ in equation~(\ref{eq:IMF}). Thus, a higher value for $m_{\rm max}$ has little effect on the results presented in this paper.

Finally, the SFR-dependent value of $\alpha_{\rm eff}$ is given as 
\begin{equation}
\label{eq:alphaeff21}
\alpha_{\rm eff}=-1.2250\log_{10}\left(\frac{SFR}{{\rm M}_{\odot}{\rm yr}^{-1}}\right) -1.4558
\end{equation}
if $\log_{10}(SFR/{\rm M}_{\odot}{\rm yr}^{-1}) \le -3.9110$,
\begin{equation}
\label{eq:alphaeff22}
\alpha_{\rm eff}=-0.23859 \log_{10}\left(\frac{SFR}{{\rm M}_{\odot}{\rm yr}^{-1}}\right) + 2.4021
\end{equation}
if $-3.9110 < \log_{10}(SFR/{\rm M}_{\odot}{\rm yr}^{-1}) \le 3.3900$, and
\begin{equation}
\label{eq:alphaeff23}
\alpha_{\rm eff}=-0.05060 \log_{10}\left(\frac{SFR}{{\rm M}_{\odot}{\rm yr}^{-1}}\right) + 1.7648
\end{equation}
if $\log_{10}(SFR/{\rm M}_{\odot}{\rm yr}^{-1}) > 3.3900$. We do follow this parametrization, but in practice, the effective high-mass slopes of almost all ETGs in the sample discussed here are given by equation~\ref{eq:alphaeff22} (see also section~3.2 in \citealt{Dabringhausen2019}).

\subsection{Linking the IGIMF to observed parameters of early-type galaxies}
\label{sec:linking}

While the SFR correlates with the shape of the IGIMF in galaxies, it is not the most practical parameter for estimating a characteristic IGIMF in early-type galaxies (ETGs) from directly observable quantities. The reason is that the present-day SFR in ETGs is so low that the present-day masses of their stellar populations imply that the SFR must have been much higher when the majority of their stars have formed. However, the abundance of $\alpha$-elements can serve as an indirect indicator for the SFR of an ETG in the past. The underlying notion is that different types of supernovae reinserted different mixtures of elements into the interstellar medium (ISM) on different timescales, from which new stars were still forming. Type-II supernovae, which have high yields of $\alpha$-elements, are thought to be the final stage of the evolution of high-mass stars and therefore occur on a timescale of Myr after the formation of their progenitor stars. Type-Ia supernovae on the other hand, which have high yields of iron, are thought to be white dwarfs that surpass the Chandrasekhar-mass by accreting matter. Depending on the parameters of the progenitor binary system, it can take individual binaries Gyrs until one of the components becomes a SNIa, while the peak in the SNIa rate in a typical ETG is at a timescale of approximately 0.3 Gyr according to \citet{Matteucci2001}. The key point is that the delay for SNIas in a galaxy is in any case longer than the delay for SNIIs. This is because SNIIs are linked to the timescale for the evolution of massive stars while SNIas require the remnant of an intermediate or low mass star. As a consequence, the SNIIs associated to a given star formation event naturally precede the SNIas associated to the same event. Long-lasting star formation leads however to a continuing production of both SNII and SNIa progenitors, and after a while newly forming stars can use preprocessed material from SNIa and SNII alike. This suggest that the higher the $\alpha$-abundance in a stellar population of a ETG is in comparison to its iron-abundance, the sooner the ETG must have stopped to form stars.

On this basis, \citet{Yan2021} have estimated the timescales for the formation of ETGs of different mass. They find
\begin{equation}
	\label{eq:deltatMdyn}
	\frac{\tau }{{\rm Gyr}} = \left\{
	\begin{matrix}
		0.003 \, \left( \frac{M_{\rm IGIMF}}{{\rm M}_{\odot}} \right) ^{0.3}   & \mbox{for} \ \frac {M_{\rm IGIMF}}{{\rm M}_{\odot}} \le 5 \cdot 10^9 \\
		49 \, \left( \frac{M_{\rm IGIMF}}{{\rm M}_{\odot}}\right)^{-0.14} & \mbox{for} \ \frac{M_{\rm IGIMF}}{{\rm M}_{\odot}} > 5 \cdot 10^9
	\end{matrix}
	\right.
\end{equation}
where $\tau$ is an estimate for the total time that it takes the galaxy form most of its stars and and $M_{\rm IGIMF}$ is an estimate for the stellar mass of an ETG according to the IGIMF-model. Note that stellar remnants are included in $M_{\rm IGIMF}$. How $M_{\rm IGIMF}$ is calculated in this paper, including $M_{\rm can}$ as a special case for when the IMF is canonical, is described  in Section~\ref{sec:masses-ETGs}.

For ETGs with $M_{\rm IGIMF}/{\rm M}_{\odot} \le 5 \cdot 10^9$, equation~\ref{eq:deltatMdyn} is a fit to the most probable values for the star formation times according to {\sc GalIMF}\footnote{https://github.com/Azeret/galIMF}, a program which combines the IGIMF theory with a model for the chemical evolution of galaxies \citep{Yan2019theory}.

In contrast to ETGs with $M_{\rm IGIMF}/{\rm M}_{\odot} \le 5 \cdot 10^9$, several attempts have been made for ETGs with $M_{\rm IGIMF}/{\rm M}_{\odot} > 5 \cdot 10^9$ to relate their star formation times to their mass (e.g. \citealt{Thomas2005,DeLaRosa2011,McDermid2015}). The best fit to the results from {\sc GalIMF} proves to be the relation from \citet{McDermid2015}, if their equation~3 is combined with equation~3 in \citet{Thomas2005} to translate [$\alpha$/Fe] into masses. This relation, and not some new fit from {\sc GalIMF} to data, is taken for equation~\ref{eq:deltatMdyn} for galaxy masses $M_{\rm IGIMF}/{\rm M}_{\odot} > 5 \cdot 10^9$ (see figure~7 in \citealt{Yan2021}).

Thus, for high-mass ETGs, the masses are stellar masses (including stellar remnants) and the IMF is canonical, because equation~3 in \citet{Thomas2005} is given in these quantities as well. In other words, the mass entering the calculations of their star formation times is $M_{\rm can}$. However, it has been argued in \citet{Yan2019results} that $M_{\rm can}$ can be exchanged with $M_{\rm IGIMF}$ without making grave errors, and indeed, replacing $M_{\rm can}$ with $M_{\rm IGIMF}$ in the figures in this paper has only minor effects. However, we carefully distinguish between $M_{\rm can}$ and $M_{\rm IGIMF}$ for all other instances in this paper.

The values for the star formation times are nonetheless only indicative anyway. Consider e.g. galaxy mergers, which are known to happen in the Universe (e.g. \citealt{Toomre1977}). If two galaxies with ongoing star formation collide, the collision is thought to provoke a starburst, which is at least qualitatively consistent with the notion that more massive galaxies form stars more rapidly than lighter galaxies. Now on the other hand, imagine two galaxies that merge after each of them has finished its star formation. According to the picture here, their formation time scales were those of the lighter progenitors, but its mass would be that of the more massive merger remnant. In the end, the time scales for galaxy formation cited here are just statistical numbers that indicate how an average ETG of a certain mass is supposed to evolve, while individual ETGs of that mass can deviate from this value quite strongly.

\subsection{The masses of ETGs according to the IGIMF-model}
\label{sec:masses-ETGs}

Estimates for the stellar mass of an ETG according to the IGIMF-model, $M_{\rm IGIMF}$, can be obtained based on existing estimates for $M_{\rm can}$, i.e. mass estimates that are based on equation~(\ref{eq:IMF}), and estimates for the masses of evolved SSPs that formed with the IGIMF. The latter are given as
\begin{equation}
M_{\rm pop}(\alpha_{\rm eff})=k_{\rm L}\int^{m_{\rm max}}_{0.1} m_{\rm rem}(m) \xi_{\rm IGIMF}(m) \, dm,
\label{eqMrem2}
\end{equation}
i.e. by integrating the present-day masses of stars and stellar remnants over the whole range of initial stellar masses. In equation~\ref{eqMrem2}, $\xi_{\rm IGIMF}(m)$ is the IGIMF given by equation~\ref{eq:IGIMF3}, $M_{\rm pop}$ is the mass of the stellar population in dependency of $\alpha_{\rm eff}$ in equation~\ref{eq:IGIMF3}, $m_{\rm max}$ is the mass of the most massive stars forming in the galaxy and is given by equation~(\ref{eq:interpolationMmax}) for $\log_{10}(SFR/{\rm M}_{\odot}{\rm yr}^{-1})<-0.878$ or $m_{\rm max}=150 \ {\rm M}_{\odot}$ for $\log_{10}(SFR/{\rm M}_{\odot}{\rm yr}^{-1}) \ge -0.878$, and $m_{\rm rem}(m)$ is the initial-to-final mass function, which expresses the masses of stellar remnants as a function of the initial mass. The factor $k_{\rm L}$ is a scaling factor that ensures that the total luminosity of the stellar population below masses of $1\, {\rm M}_{\odot}$ remains constant when $\alpha_{\rm eff}$ varies. This corresponds approximately to the stars that still shine in old populations like the ETGs. This is motivated by the notion that the luminosity is fixed through observations, while the mass of the stellar population is treated here as an unknown parameter that is to be determined. This stands in contrast to $k$ in equation~\ref{eq:IGIMF3}, which keeps the integral at unity, independent of the value for $\alpha_{\rm eff}$. Thus, $k$ has to change with $\alpha_{\rm eff}$ in equation~\ref{eq:IGIMF3}.

The initial-to-final mass function used in equation~(\ref{eqMrem2}) is the one introduced in \citet{Dabringhausen2009}, which is designed for stellar systems older than $10^8$ years. It is given as
\begin{equation}
\label{eq:mrem}
\frac{m_{\rm rem}}{{\rm M}_{\odot}}=
\begin{cases}
\frac{m}{{\rm M}_{\odot}} & {\rm if \ \ } \frac{m}{{\rm M}_{\odot}} < \frac{m_{\rm to}}{{\rm M}_{\odot}}\\
0.109 \, \frac{m}{{\rm M}_{\odot}}+0.394 & {\rm if \ \ } \frac{m_{\rm to}}{{\rm M}_{\odot}} \le \frac{m}{{\rm M}_{\odot}} < 8\\
1.35 & {\rm if \ \ } 8  \le  \frac{m}{{\rm M}_{\odot}} < 25\\
a \, \frac{m}{{\rm M}_{\odot}} & {\rm if \ \ } 25 \leq \frac{m}{{\rm M}_{\odot}} \le m_{{\rm max}}.
\end{cases}
\end{equation}
$m_{\rm max}$ is given in this equation by equation~(\ref{eq:interpolationMmax}) for $\log_{10} (SFR/{\rm M}_{\odot}{\rm yr}^{-1})< -0.878$ and else by $150 \, {\rm M}_{\odot}$.

Thus, stars with masses below the main-sequence turn-off mass $m_{\rm to}$ are considered to still have their initial masses, white dwarfs are thought to have progenitors with masses between $m_{\rm to}$ and $8 \, {\rm M}_{\odot}$ and their masses are given by a relation found by \citet{Kalirai2008} and neutron stars are thought to have progenitors with masses between $8 \, {\rm M}_{\odot}$ and $25 \, {\rm M}_{\odot}$ and are all considered to have a mass of $1.35 \, {\rm M}_{\odot}$, which is observationally supported by \citet{Thorsett1999}. Stars with initial masses above $25 \, {\rm M}_{\odot}$ are considered to evolve into black holes. The mass of these black holes is the most uncertain parameter and strongly depends on their metallicity (compare for example figures~12 and~16 in \citealt{Woosley2002}). In equation \ref{eq:mrem}, it is simply set to a constant $a$, $0\le a <1$. Thus, $a$ is a factor anywhere in between the complete destruction of the star, or the almost complete perseveration of the mass of the star in the process of becoming a BH. The case $a=1$ is impossible, because stars produce winds during their lifetimes. On the other hand, the existence of stellar-mass BHs is well known, either though the detection of X-rays that such a BH emits if it accretes matter (e.g. \citealt{Clark1975} and \citealt{Ivanova2008}), or through the direct detection of gravitational waves in mergers of stellar-mass BHs \citep{Abbott2016}.

We reflect this uncertainty on the mass of the BHs by considering the two cases, which mark the possible extremes to the values that can be obtained.

In the first case, we use the assertion that low BH masses correspond to stars with high metallicities, and high black hole masses to stars with low metallicities. This progression of the mass of the black holes with metallicity is estimated here as
\begin{equation}
\label{eq:blackhole}
a=\frac{m_{\rm BH}}{m}=-\frac{4}{3} [Z/{\rm H}] + \frac{1}{6},
\end{equation}
where $m_{\rm BH}$ is the mass of the black hole, $[Z/{\rm H}]$ is the metallicity of the galaxy today and $m$ is the initial mass of the progenitor star. Equation~\ref{eq:blackhole} produces black holes that have 0.5 times the mass of the progenitor at $[Z/{\rm H}]=-2.5$ and decline linearly to 0.1 times the progenitor mass at $[Z/{\rm H}]=0.5$. By using the contemporary metallicities of the galaxies, the masses of the BHs are especially for the metal-rich massive ETGs likely to be too low.

In the second case, we maximize the mass of the BHs by setting $a=1$, i.e. the massive stars have not experienced any mass loss before they evolve to BHs. We do this despite the fact that every star produces a stellar wind, albeit the metal-poor ones much less than the metal-rich ones (e.g. \citealt{Woosley2002}).

The explicit terms that the integration in equation~(\ref{eqMrem2}) yields are listed in the appendix to \citet{Dabringhausen2009}, provided that 0.1 in their last equation is replaced by $-4/3[Z/{\rm H}]+1/6$ for the first case and 1 for the second case.

The estimates for the actual masses of the stellar populations of the ETGs according to the IGIMF-model can then be calculated as
\begin{equation}
M_{\rm IGIMF} = \left(\frac{M_{\rm pop}(\alpha_{\rm eff})}{M_{\rm pop}(\alpha_{\rm eff}=2.3)}\right)\times M_{\rm can},
\label{eq:MIGIMF}
\end{equation}
where $M_{\rm pop}(\alpha_{\rm eff})$ is given by evaluating equation~(\ref{eqMrem2}) for the $\alpha_{\rm eff}$ implied by the SFR with which the majority of the stellar population of the studied ETG has formed and $m_{\rm max}$ is given by equation~(\ref{eq:interpolationMmax}). $M_{\rm pop}(\alpha_{\rm eff}=2.3)$ is given by evaluating equation~(\ref{eqMrem2}) for $\alpha_{\rm eff}=2.3$, and $M_{\rm can}$ is the estimate of the mass that the stellar population of the studied ETG would have with $\alpha_{\rm eff}=2.3$. Note that the results of equation~\ref{eq:MIGIMF} can change even for the same value of $\alpha_{\rm eff}$, if the assumed masses for the stellar-mass BHs change.

As the stellar populations in question are typically much older than one Gyr, all stars that have survived until now have masses around 1$\, {\rm M}_{\odot}$, or less. For such stars, the precise ages of the stars do not matter much for the luminosity of their stellar population, if the stellar population follows eq.~\ref{eq:IGIMF3}. However, the value of the high-mass slope of the IGIMF, $\alpha_{\rm eff}$, continues to matter for the mass of the stellar population, and thus also for the mass-to-light ratio. Simply put, the same difference in $\alpha_{\rm eff}$ makes much less of a difference in the relative numbers of stars between $1 \, {\rm M}_{\odot}$ and $2 \, {\rm M}_{\odot}$ than for stars with masses between, say, $25 \, {\rm M}_{\odot}$ and $26 \, {\rm M}_{\odot}$. This is explained in detail in section 3.4 in \citet{Dabringhausen2019}.

\subsection{Dynamical masses}
\label{sec:dynmass}

Besides the stellar masses, i.e. mass estimates based on assumptions on the stellar populations of the ETGs, there are also dynamical mass estimates based on the structure and the motions in the galaxies. In its simplest form, some effective radius, a velocity dispersion within this radius, and some density profile are needed for a dynamical mass estimate. More elaborate dynamical mass models could e.g. work with several velocity dispersions from different spots of the same galaxy, or distinguish between velocity dispersions and rotational velocities of a galaxy. The availability of both mass estimates (i.e. dynamical and stellar) is crucial for testing laws of gravity, i.e. the work done here. If both agree with one another within the uncertainties, it means that both ways of measuring the mass of the galaxy yield the same result.

First, we make some assertions, assumptions and definitions on the galaxies in our sample. They are vital for the estimates of their dynamical masses:

\begin{itemize}
	
\item The mass profiles of the galaxies are S\'{e}rsic-profiles (\citealt{Sersic1963}; see section~\ref{sec:data-sersic}), as opposed for example to Plummer-profiles \citep{Plummer1911} or King-profles \citep{King1962}. The S\'{e}rsic-profiles are more complicated than in particular the Plummer profile, but they are especially adapted to galaxies, and not to star clusters.
	
\item Massive ETGs are known to possess a central peak in their $M_{\rm dyn}/L$-ratios, indicating more mass per unit luminosity at their centres \citep{vanDokkum2010}. \citet{Bernardi2018} model this as 
\begin{equation}
	\label{eq:MLpeak}
	\Upsilon=\Upsilon_{\rm can} \, [1+c\, (\epsilon - \eta \, (R / R_{\rm e}))]
\end{equation}
for $R$ where $\Upsilon > \Upsilon_{\rm can}$, and $\Upsilon = \Upsilon_{\rm can}$ otherwise. In this equation, $\Upsilon$ is the mass-to-light ratio at the projected radius $R$, $\Upsilon_{\rm can}$ is the mass-to-light ratio with the canonical IMF and $R_{\rm e}$ is the projected half-light radius. The parameter $c$ is given as
\begin{equation}
	c = \frac{[\sigma_0 / (km \, s^{-1})]-100}{250 - 100},
\end{equation}
where $\sigma_0$ is the central velocity dispersion. For the parameters $\epsilon$ and $\eta$, \citet{Bernardi2018} give two choices: Either $\epsilon=2.33$ and $\eta=6.0$ based on observation of six massive ETGs by \citet{vanDokkum2017}, or a more intermediate case with  $\epsilon=1.29$ and $\eta=3.33$. Subsequent observation of many more ETGs by \citet{Chae2018} and \citet{DominguezSanchez2019} favour the intermediate case. In this paper, this will be done as well, and referred to as the `standard case'.

If $\sigma_0 > 100 \,$km/s and the IGIMF-model is used for estimating the mass of a given ETG, we integrate 
\begin{equation}
M = 2 \pi \int_{0}^{R} I(R) \, \Upsilon \, R \, dR
\end{equation}
with increments of $dR=1 \, {\rm parsec}$, until $M/M_{\rm IGIMF}>0.5$ is reached. $I(R)$ is the S\'{e}rsic-profile as given in equation~\ref{eq:sersic-profile}, and we start with the S\'{e}rsic-index $n$ given in \citet{Dabringhausen2016a}. The corresponding value of $R$ is the new half-mass radius, which is smaller than the half-mass calculated with mass following the light of the galaxy. Also $n$ needs to be updated to the new $R_{\rm e}$ with equation~\ref{eq:sersic}. Finding the updated $R_{\rm e}$ and $n$ is in principle an iterative process, but we stop here after the first iteration.

If the IGIMF-model is not used, or $\sigma_0 < 100 \,$km/s in the given ETG, then mass follows light.

\item Gas and dust are neglected in the ETGs (see \citealt{Young2011} for the paucity of gas in ETGs and \citealt{Dariush2016} for the paucity of dust).

\item The default for the ETGs is an anisotropy parameter, $\beta$, at zero (cf. equation~4-53b in \citealt{Binney1987}). This describes a galaxy that does not rotate and whose systematic patterns in the motions of the stars can be neglected. As a consequence, the dynamical mass of the ETG will be overestimated if $-\infty<\beta<0$ (i.e. the stellar orbits in the ETG are predominately circular) and underestimated if $0<\beta<1$ (i.e. the orbits in the ETG are predominately radial). We also introduce rotation of the galaxies if some value on it was published in \citet{Dabringhausen2016a}, and refer the reader to Sections~\ref{sec:bin} to~\ref{sec:bin+rot} in the present paper for details on it.

\item The ETGs are in virial equilibrium. This means that the ETG has settled into a configuration where its gravitational potential and thus its density profile does not depend on time. This assumption implies in particular that the ETG are not significantly disturbed by tidal forces, i.e. external, time-dependent gravitational fields. Tidal fields increase the actually observed internal velocity dispersions and may even disrupt galaxies \citep{Kroupa1997,Fellhauer2006,McGaugh2010,Casas2012,Dominguez2016}. Galaxies which are wrongly assumed to be in virial equilibrium when their dynamical mass is estimated, but are in fact subjected to tides, can have dynamical ´masses' which are too high by orders of magnitudes.

\item The density profiles of the ETGs can be approximated as spherically symmetric. As a consequence of this assumption, we use in the following $R_{\rm e}$ to estimate the dynamical mass of an ETG (instead of elliptic shapes, for instance). $R_{\rm e}$ is also used for estimating the characteristic accelerations in the ETGs due to their stellar or dynamical masses.

\end{itemize}

These assumptions can be cast together into an estimate for the dynamical masses of the galaxies as
\begin{equation}
M_{\rm dyn}=\frac{K_{\rm V}}{G}\, R_{\rm e}\sigma_{0}^2,
\label{eq:Mdyn}
\end{equation}
where $K_{\rm V}$ is a factor that depends on the shape of the density profile of the ETG and $G$ is the gravitational constant. $K_{\rm V}$ is approximated with equation~(11) in \citet{Bertin2002}, i.e.
\begin{equation}
K_{\rm V}(n) = \frac{73.32}{10.465+(n-0.94)^2}+0.954,
\end{equation}
where $n$ is the S\'{e}rsic index. Thus, $K_{\rm V}$ increases with decreasing $n$, which in turn decreases with decreasing $R_{\rm e}$. If $\sigma_0$ is left unchanged, assuming a peak in $\Upsilon$ at the centre of an ETG decreases $M_{\rm dyn}$ compared to $\Upsilon$ being constant throughout the ETG.

The kinetic Energy of an ETG is, under the assumptions in this section,
\begin{equation}
\label{eq:Ekin}
E_{\rm kin}=\frac{1}{2} \, M_{\rm dyn} \, \sigma_0^2,
\end{equation}
so that 
\begin{equation}
\label{eq:altMdyn}
M_{\rm dyn}=\sqrt{\frac{2 \, K_{\rm V} \, R_{\rm e}}{G}} \, E_{\rm kin}
\end{equation}
is an alternative way to express equation~\ref{eq:Mdyn}.

The simplest way to estimate a dynamical mass with equation~\ref{eq:Mdyn} is to take an observed central line-of-sight velocity dispersion, $\sigma_{\rm obs}$, and set it equal to $\sigma_0$. However, this approach neglects the effects binaries and the rotation of the galaxies may have on the dynamical mass estimates. Thus, binaries and rotation are introduced in Sections~\ref{sec:bin} to~\ref{sec:bin+rot}.

\subsection{Binaries}
\label{sec:bin}

\begin{figure}
\centering
\includegraphics[scale=0.85]{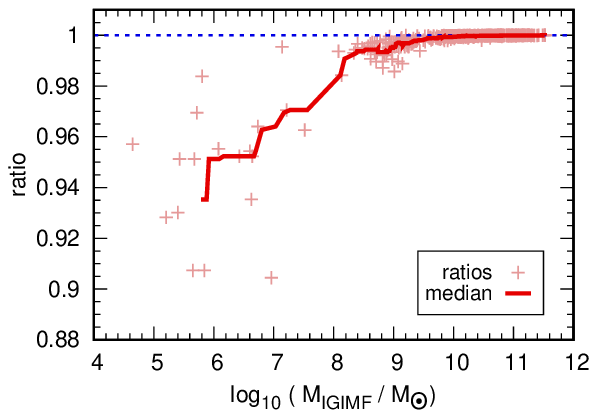}
\caption[The ratios $M_{\rm dyn}^{\rm bin}/M_{\rm dyn}$ in dependence of $M_{\rm IGIMF}$]{\label{fig:bin} The ratios $M_{\rm dyn}^{\rm bin}/M_{\rm dyn}$ in dependence of $M_{\rm IGIMF}$ for all 462 ETGs in the sample as light red crosses. The thick dark red line indicates for each galaxy the median value of itself and its 20 closest neighbours in $M_{\rm IGIMF}$. An exception are the ten least massive ETGs and the ten most massive ETGs, for which this quantity cannot be calculated, because this quantity should incorporate 10 ETGs with smaller masses and 10 ETGs with larger masses.}
\end{figure}

Consider the presence of binaries in the galaxies. The motion of a star in a binary has two components, namely the motion around the centre of mass of the binary and the motion of the centre of mass in the potential of the ETG. The distribution of the line-of-sight motions of the stars in binaries are the consequence of random combinations of many parameters and processes (such as direction motion in 3D-space, orientation of binaries in 3D-space, excentricities, energy exchange in encounters, and so on). Thus, the central limit theorem implies that this distribution is Gaussian. The motions of stars in binaries and the motions of the stellar systems  (i.e. single stars and binary systems) in the ETG are moreover uncorrelated. This means that the observed line-of-sight velocity dispersion of a stellar population is given as
\begin{equation}
\label{eq:binaries}
\sigma_{\rm obs}^2 =(1-f)\sigma_0^2+f(\sigma_0^2+\sigma_{\rm bin}^2),
\end{equation}
where $\sigma_{\rm obs}$ is the {\it observed} central line-of-sight velocity dispersion, $\sigma_0$ is the central line-of-sight velocity dispersion of the tracer population due to the {\it potential} of the ETG and $\sigma_{\rm bin}$ is the line-of-sight velocity dispersion of stars in {\it binaries} due to their motions around their common center of mass, $f$ is the fraction of binaries and thus $1-f$ is the fraction of singles. In principle, also hierarchical higher-order multiples would have to be considered, if present-day stellar populations are dealt with. Their number may seem significant if compared to binaries (see e.g. \citealt{Raghavan2010}), but in fact every hierarchical multiple system consists of one or more binaries at the bottom of the hierarchy. These binaries are the tightest units of the complete system and their star move thus the fastest around their centres of masses. They would therefore completely dominate the increase of $\sigma_{\rm obs}$ over $\sigma_0$. Hence, higher order multiples are neglected here, and
\begin{equation}
M_{\rm dyn}^{\rm bin}=\frac{K_{\rm V}}{G}R_{\rm e} \sigma_0^2 = \frac{K_{\rm V}}{G}R_{\rm e} (\sigma_{\rm obs}^2- f \sigma_{\rm bin}^2).
\label{eq:Mdyn-bin}
\end{equation}
is the true dynamical mass of an ETG with binaries under this premise.

According to \citet{Marks2011field}, the binary fraction $f$ is about 80 percent in very low-mass ETGs, whereas it is about 30 percent in high-mass ETGs. We emulate this dependency by assuming $f=0.8$ for ETGs with $M_{\rm can} = 10^3 \, {\rm M}_{\odot}$, and then linearly going down with $M_{\rm can}$, until we reach $f=0.3$ for ETGs with $M_{\rm can} = 10^{12} \, {\rm M}_{\odot}$. For the value of $\sigma_{\rm bin}$, \citet{Gieles2010} derive an analytic approximation, which they formulate as
\begin{equation}
\sigma_{\rm bin}^2=\frac{1}{3}\left(\frac{2q^{3/2}}{1+q}\right)^{4/3} \left(\frac{\pi G m_1}{2P}\right)^{2/3}.
\label{eq:sigma-bin}
\end{equation}
In this equation, $q$ is the mass ratio of the two components of the binary, $m_1$ is the typical mass of the more massive component of the binary and $P$ is the typical orbital period in the binaries. In this paper, we use $q=0.6$ (cf. \citealt{Gieles2010}), $P=10$ yr and $m_1 = 1 \ {\rm M}_{\odot}$, which leads to $\sigma_{\rm bin} = 2.3$ km/s in equation~(\ref{eq:sigma-bin}); see \citealt{Dabringhausen2016b}. Thus, in comparison with the median in $P$ of 180 yr in \citet{Duquennoy1991}, the binaries are very tight. We therefore maximize their effect on $\sigma_{\rm bin}$, and thus on $\sigma_{0}$, which together have to result in the observable $\sigma_{\rm obs}$. Choosing the binaries very tight probably also overcompensates the small error made by ignoring the higher orders of hierarchical stellar systems with more than two stars.

Fig.~\ref{fig:bin} shows the ratios of the dynamical masses with and without binaries, plotted over the stellar mass of their galaxies according to the IGIMF-model. Especially at low masses, $M_{\rm dyn}^{\rm bin}$ is lower than $M_{\rm dyn}$ in selected galaxies, but at high stellar masses, both ways to esstimate the mass yield almost the same results. Note that this is not an effect of the more massive galaxies having fewer binaries (see \citealt{Marks2011field}), but of the more massive galaxies having $\sigma_0$ so large that $\sigma_{\rm bin}$ becomes negligible. This is also the case for our rather generous choice of $\sigma_{\rm bin}$.

We calculate also a median of the ratios of $M_{\rm dyn}^{\rm bin}$ in dependence of $M_{\rm dyn}$, which is shown as a solid (red) line in Fig.~\ref{fig:bin}. For obtaining these medians, we consider all ETGs for which the ratios of the mass estimates are known, and number them consecutively by ascending stellar mass. From this list, we construct all possible subsets that contain 21 ETGs with consecutive numbers in stellar mass. These subsets are then sorted with ascending ratios of $M_{\rm dyn}^{\rm bin}$ over $M_{\rm dyn}$, and the 11th highest value among this subset is then chosen as the characteristic value for the ratio of ETGs with the luminosity of the $i$th ETG. By construction, this method cannot assign a characteristic value for $M_{\rm dyn}^{\rm bin}/M_{\rm dyn}$ to the 10 least least luminous and the 10 most luminous ETGs. However, the advantage of this method is that the impact of ETGs with exceptional $M_{\rm dyn}^{\rm bin}/M_{\rm dyn}$ on the estimates of the typical $M_{\rm dyn}^{\rm bin}/M_{\rm dyn}$ is minimized.

Note that the effect of binaries on the galactic dynamics is larger in \citet{Dabringhausen2016b} than here; especially for low-mass ETGs. The reason for this is that in \citet{Dabringhausen2016b}, the dynamics predicted by the amount of baryons are taken as the basis, and these defaults are increased by the expected influence that binaries may have. This effect is quite substantial especially for the case of Newtonian dynamics. The least luminous ETGs are expected to have internal velocity dispersions well below 1 km/s in this case, which is by about an order of magnitude lower than $\sigma_{\rm bin}=2.3$ km/s. In this paper in contrast, the observed internal velocity dispersions are considered as the default values, and then $\sigma_{\rm bin}$ is substracted. The conclusion of \citet{Dabringhausen2016b} however, namely that binaries do not increase the dynamical mass enough to substitute the cold dark matter that the ETGs would have according to the $\Lambda$CDM-model, becomes even stronger with the approach taken here.

\subsection{Rotation}
\label{sec:rot}

\begin{figure}
\centering
\includegraphics[scale=0.85]{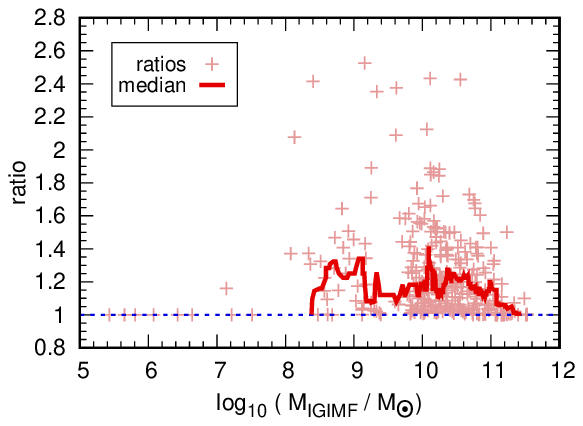}
\caption[The ratios $M_{\rm dyn}^{\rm rot}/M_{\rm dyn}$ in dependence of $M_{\rm IGIMF}$]{\label{fig:rot} The ratios $M_{\rm dyn}^{\rm rot}/M_{\rm dyn}$ in dependence of $M_{\rm IGIMF}$ for the 310 galaxies of the 462 ETGs in the full sample, for which some estimate of $v_{\rm rot}$ is available. They are shown as light red crosses. The thick dark red line indicates for each galaxy the median value of itself and its 20 closest neighbours in $M_{\rm IGIMF}$. An Exception are the ten least massive ETGs and the ten most massive ETGs, for which this quantity cannot be calculated, because this quantity should incorporate 10 ETGs with smaller masses and 10 ETGs with larger masses.}
\end{figure}

Consider the case of an ETG without binaries, but an average rotation $v_{\rm rot}$. We furthermore assume that $v_{\rm rot}$ (i.e ordered motion) and $\sigma_0$ (i.e. random motion) are the means of symmetric functions (e.g. Gaussian). The total kinetic energy of the two motions can then be summed up as
\begin{equation}
\label{eq:Trot}
E_{\rm kin}=\frac{1}{2}M_{\rm dyn}  \langle v^2 \rangle = \frac{1}{2}M_{\rm dyn}(\sigma_0^2+v_{\rm rot}^2),
\end{equation}
where $\langle v^2 \rangle$ is the summation over all squared velocities. Thus, in analogy to equation~\ref{eq:altMdyn},
\begin{equation}
\label{eq:Mdyn-rot}
M_{\rm dyn}^{\rm rot}=\sqrt{\frac{2 \, K_{\rm V} \, R_{\rm e}}{G}} E_{\rm kin}=\frac{K_{\rm V}}{G}\, R_{\rm e}(\sigma_0^2+v_{\rm rot}^2)
\end{equation}
holds for the rotating ETGs.

Fig.~\ref{fig:rot} shows the ratios of the dynamical masses with and without rotation, i.e. $M_{\rm dyn}^{\rm rot}$ over $M_{\rm dyn}$, plotted over the stellar mass of their galaxies according to the IGIMF-model. The dynamical masses are often substantially higher if also the rotation of the ETGs is considered; sometimes more than twice as high. Only at very low stellar masses, the two estimates for the dynamical mass are often the same, but this could also be because rotations are very hard to detect at their luminosities. Also, recall that the estimates of $v_{\rm rot}$ are pretty basic in some ETGs, as described in \citet{Dabringhausen2016a} and Section~\ref{sec:data-rotation}.

Also a median, analogously to Fig~\ref{fig:bin}, was calculated in Fig.~\ref{fig:rot}. This median starts at unity for very low stellar masses (see above), and ends at unity for the ETGs with the highest stellar masses, but oscillates from $M_{\rm dyn}^{\rm rot}/M_{\rm dyn} \approx 1.1$ to $ M_{\rm dyn}^{\rm rot}/M_{\rm dyn} \approx 1.4$ in between.

Hence, a neglect of the rotation leads to smaller dynamical masses in rotating systems, and thus opposed to the effect of binaries.

\subsection{Binaries and rotation}
\label{sec:bin+rot}

Finally, the effects of binaries and rotation on the dynamical masses of ETGs can also be combined to 
\begin{equation}
\label{eq:Mdyn-binrot}
M_{\rm dyn}^{\rm bin+rot}=\frac{K_{\rm V}}{G}\, R_{\rm e}(\sigma_{\rm obs}^2-f\sigma_{\rm bin}^2+v_{\rm rot}^2).
\end{equation}
However, a plot of the ratios of $M_{\rm dyn}^{\rm bin+rot}$ over $M_{\rm dyn}$ is next to identical to a plot of $M_{\rm dyn}^{\rm rot}$ over $M_{\rm dyn}$, i.e. Fig.~\ref{fig:rot}. We nevertheless compare $M_{\rm dyn}^{\rm bin+rot}$ rather than $M_{\rm dyn}^{\rm rot}$ to $M_{\rm can}$, and $M_{\rm IGIMF}$, respectively, in Section~\ref{sec:results}, simply because $M_{\rm dyn}^{\rm bin+rot}$ is more comprehensive than $M_{\rm dyn}^{\rm rot}$.

\subsection{Radial accelerations}
\label{sec:accelerations}

\begin{figure*}
\centering
\includegraphics[scale=0.85]{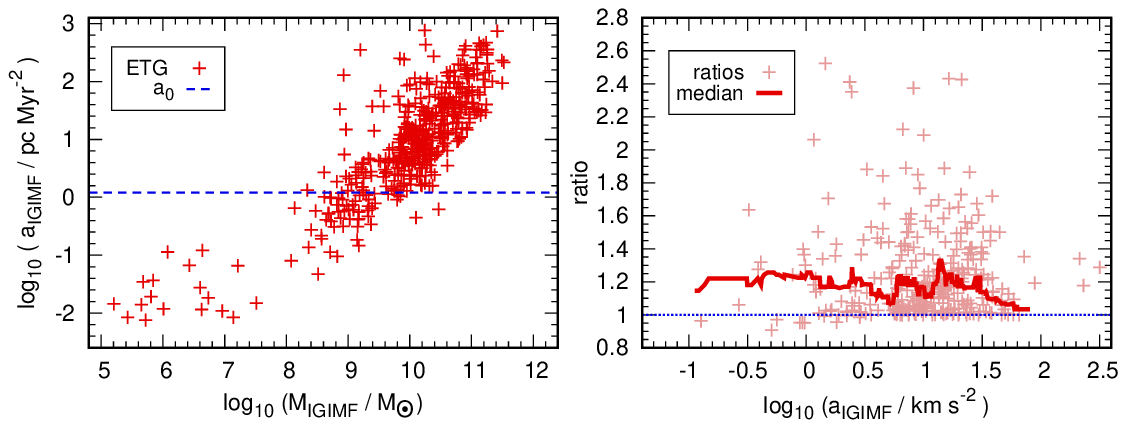}
\caption[The ratios $M_{\rm dyn}^{\rm bin+rot}/M_{\rm dyn}$ over $a_{\rm IGIMF}$]{\label{fig:acc-mass} Left panel: The stellar radial accelerations, $a_{\rm IGIMF}$, in dependence of the stellar masses, $M_{\rm IGIMF}$, of the 462 ETGs in the full sample calculated with equation~\ref{eq:IGIMF3}. High values for $a_{\rm IGIMF}$ are strongly correlated to high values in $M_{\rm IGIMF}$. Right panel: The ratios $M_{\rm dyn}^{\rm bin+rot}/M_{\rm dyn}$ in dependence of $a_{\rm IGIMF}$ for the 310 galaxies of the 462 ETGs in the full sample, for which some estimate of $v_{\rm rot}$ is available. They are shown as light red crosses. Note that $M_{\rm dyn}^{\rm bin+rot}$ is almost equal to $M_{\rm dyn}^{\rm rot}$ on the y-axis, and thus a comparison to Fig.~\ref{fig:bin} essentially only shows the rearrangement of the data by switching from $M_{\rm IGIMF}$ to $a_{\rm IGIMF}$ on the x-axis. The thick dark red line indicates for each ETG the median value of itself and its 20 closest neighbours in $a_{\rm IGIMF}$. An exception are the ten ETGs with the lowest values in $a_{\rm IGIMF}$ and the ten ETGs with the highest values in $a_{\rm IGIMF}$, for which this quantity cannot be calculated, because this quantity should incorporate 10 ETGs with smaller radial accelerations and 10 ETGs with larger radial accelerations. In both panels, the mass of the stellar-mass BHs is based on equation~\ref{eq:blackhole} and the metallicity the ETGs have today.}
\end{figure*}

When discussing theories of gravity, it is sometimes useful to express the dynamical radial accelerations, $a_{\rm dyn}$ as functions of the prediction for the radial acceleration based on the amount of matter inside the enclosed volume. In Newtonian dynamics and with the assumptions made on ETGs made in Section~\ref{sec:dynmass}, this is simply
\begin{equation}
a_{\rm dyn}^{\rm Newt}=a_{\rm star},
\label{eq:RAR-newton}
\end{equation}
where the left side of the equation is the dynamical acceleration in Newtonian dynamics, and the right side of the equation is the acceleration due to the stars inside the radius where the dynamical acceleration was measured. The subscript `star' on the left side of the equation signifies that the stellar mass could come from a stellar population with any mass spectrum. Equation~\ref{eq:RAR-newton} is thus the radial acceleration relation (RAR) for Newtonian gravity. It will be referred to as the Newtonian RAR.
 
The dependency between the two radial accelerations can also be described by the $\mu$-function, which is in the case of Newtonian dynamics given as
\begin{equation}
\mu\left(\frac{a_{\rm dyn}}{a_0}\right)\times a_{\rm dyn} = a_{\rm star},
\label{eq:MOND}
\end{equation}
where $a_0$ is Milgrom's constant. This implies
\begin{equation}
\label{eq:mu-newton}
\mu \left( \frac{a_{\rm dyn}^{\rm Newt}}{a_0} \right)=1.
\end{equation}
for the $\mu$-function in Newtonian dynamics. 

Things become more interesting in Milgromian dynamics, or MONDian dynamics, where $a_0$ is the acceleration at the transition between the quasi-Newtonian regime and the deeply Milgromian regime. It is a fundamental constant in Milgromian dynamics, which we set to $1.2\times 10^{-10} \rm{m \, s^{-2}}$, in concordance with \citet{Famaey2005}.

In this paper, we pick two $\mu$-functions for Milgromian dynamics, because they are simple, and widely used to date.
	
The first Milgromian $\mu$-function is the standard $\mu$-function, which was introduced by \citet{Milgrom1983gal} as
\begin{equation}
\mu\left(\frac{a_{\rm dyn}^{\rm std}}{a_0}\right)=\frac{a_{\rm dyn}^{\rm std}}{a_0}\left[1+\left(\frac{a_{\rm dyn}^{\rm std}}{a_0}\right)^2\right]^{-\frac{1}{2}},
\label{eq:mu-standard}
\end{equation}
and implies, in combination with equation~(\ref{eq:MOND}) and solving it for $a_{\rm dyn}^{\rm std}$,
\begin{equation}
a_{\rm dyn}^{\rm std}=a_{\rm star}\left(\frac{1}{2}+\frac{1}{2}\left[1+\left(\frac{2 \, a_0}{a_{\rm star}}\right)^2\right]^{\frac{1}{2}}\right)^{\frac{1}{2}}
\label{eq:RAR-standard}
\end{equation}
for the RAR. This theoretical RAR will be referred to as the standard-MOND RAR.

Another possible choice for the $\mu$-function is, according to \citet{Famaey2005} and \citet{Zhao2006}, the simple $\mu$-function, which is given as
\begin{equation}
\mu\left(\frac{a_{\rm dyn}^{\rm spl}}{a_0}\right)=\frac{a_{\rm dyn}^{\rm spl}}{a_0}\left[1+\left(\frac{a_{\rm dyn}^{\rm spl}}{a_0}\right)\right]^{-1}
\label{eq:mu-simple}
\end{equation}
and implies in combination with equation~(\ref{eq:MOND})
\begin{equation}
a_{\rm dyn}^{\rm spl}=a_{\rm star}\left[\frac{1}{2}+\frac{1}{2}\left(1 +\frac{4 \, a_0}{a_{\rm star}}\right)^{\frac{1}{2}}\right]
\label{eq:RAR-simple}
\end{equation}
for the RAR. This theoretical RAR will be referred to as the simple-MOND RAR.

Note that high-precision measurements at the accelerations in the Solar system formally exclude both functions to different degrees \citep{Blanchet2011,Hees2016}. However, for the typical accelerations at the half-mass radii of galaxies, which are much lower, they are both fine, with a possible lead for the simple $\mu$-function (e.g. \citealt{Famaey2005,Gentile2011,Milgrom2012,Janz2016,Chae2019radial}). 
	
There are many more subtly different shapes for $\mu(a_{\rm dyn}/a_0)$ discussed in the literature, since the correct formulation of $\mu(a_{\rm dyn}/a_0)$ does to date not follow from Milgromian dynamics itself. For instance, the $\mu$-function introduced by \citet{McGaugh2008} is consistent with the observations in the Solar system, but is difficult to solve for $a$. However, our main concern here is that the $\mu$-function correctly describes the data in the range relevant to this paper, where the $\mu$-function by \citet{McGaugh2008} closely resembles the simple $\mu$-function. Thus, nothing new would be added by considering it as well.
	
The next step is to compare these theoretical RARs with observational accelerations in galaxies, as discussed by e.g. \citet{Wu2015}, \citet{McGaugh2016} or \citet{Lelli2017}. The mass of the galaxy inside the radius where the observation took place, i.e. where the matter responsible for its dynamics is contained, is at least to some extent based on models and assuptions (such as virial equilibrium, the mass-to-light ratio of the stellar population, the presence of a CDM-halo, and so on). If the underlying model of a given theoretical RAR fully accounts for the properties of the considered ETGs, the data on the ETGs will be located along the respective theoretical RAR. However, agreement between the data and the theoretical RAR is not a sufficient condition for the underlying assumptions to be correct, while a systematic deviation from the theoretical RAR indicates for a certainty that some of the made assumptions are wrong, or some important aspect is missing in the models for the ETGs.

In principle, a RAR can be tested on many different radii of the same galaxy. However, here we test every galaxy only once at its half-mass radius $R_{\rm h}$, which is derived according to the assumptions made in Section~\ref{sec:dynmass}. Moreover, according to the appendix in \citet{Wolf2010}, $R_{\rm h} \approx (4/3) \, R_{\rm e}$. This is remarkably independent of the mass profile actually used to model the ETG or the star cluster. Thus, 
\begin{equation}
\label{eq:acc-obs}
a_{\rm dyn}^{\rm obs}=\frac{1}{2}\frac{G \, M_{\rm dyn}^{\rm obs}}{R_{\rm h}^2} \approx \frac{9}{32}\frac{G \, M_{\rm dyn}^{\rm obs}}{R_{\rm e}^2}
\end{equation}
for the observed radial acceleration. 

The corresponding predicted stellar radial acceleration to $a_{\rm dyn}^{\rm obs}$ is in this paper  
\begin{equation}
\label{acc-rh}
a_{\rm IGIMF}=\frac{1}{2}\frac{G \, M_{\rm IGIMF}}{R_{\rm h}^2} \approx \frac{9}{32}\frac{G \, M_{\rm IGIMF}}{R_{\rm e}^2},
\end{equation}
where $M_{\rm IGIMF}$ is the stellar mass . Multiplication of equation~\ref{acc-rh} with the according $\mu$-function would give the prediction of the dynamical radial acceleration at this stellar radial acceleration; e.g. equations~\ref{eq:mu-newton},~\ref{eq:mu-standard} and~\ref{eq:mu-simple} for the choices in this paper. 

The dependency between stellar radial relations, $a_{\rm IGIMF}$, and stellar mass, $M_{\rm IGIMF}$, is illustrated in Fig.~\ref{fig:acc-mass} for the specific case of the simplified IGIMF (equation~\ref{eq:IGIMF3}). In the left panel of Fig.~\ref{fig:acc-mass}, high values in $a_{\rm IGIMF}$ are clearly strongly correlated with high values in $M_{\rm IGIMF}$. This is not obvious. Imagine two ETGs, which have the same stellar mass, but different half-mass radii. The radius shall be much larger in the second ETG than in the first ETG, so that the first ETG will have a larger stellar radial acceleration at its half-mass radius. However, in reality, the rise of the average radii with luminosity, or mass (see figure~2 in \citealt{Dabringhausen2016a}) is not strong enough to negate or reverse the trend of rising accelerations with rising luminosity, or mass. In the right panel of Fig.~\ref{fig:acc-mass}, the ratios of $M_{\rm dyn}^{\rm bin+rot}/M_{\rm dyn}$ are plotted over their stellar radial accelerations. Also the medians of $M_{\rm dyn}^{\rm bin+rot}/M_{\rm dyn}$ are shown as a (red) solid line. They are created as in Fig.~\ref{fig:bin}, but based on the stellar radial accelerations instead of their stellar masses. The right panel of Fig.~\ref{fig:acc-mass} can thus be compared to Fig.~\ref{fig:rot}, which shows essentially the same data on the y-axis (see the discussion on $M_{\rm dyn}^{\rm rot}$ and $M_{\rm dyn}^{\rm rot+bin}$ in Section~\ref{sec:bin+rot} of the present paper), but with stellar masses instead of the stellar radial accelerations on the x-axis.

\section[Results]{Results}
\label{sec:results}

\subsection{ETGs with the canonical IMF}
\label{sec:canonical}

\begin{figure*}
\centering
\includegraphics[scale=0.88]{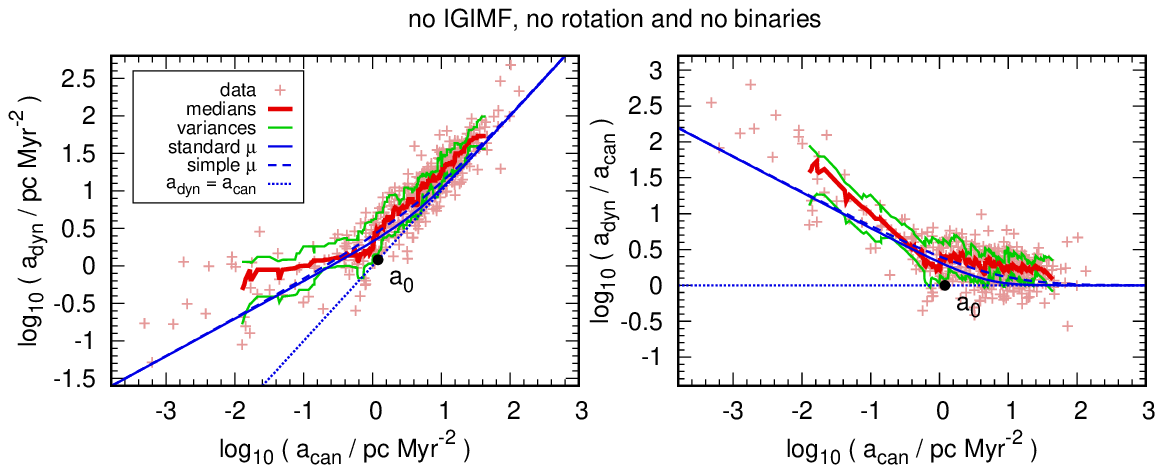}
\caption[The dynamical radial accelerations for ETGs over the expected acceleration based on $M_{\rm can}$ 1]{\label{fig:can-norot} Left panel: The radial accelerations for ETGs without accounting for the IGIMF. On the x-axis is the expected acceleration based on the amount of baryons present in the ETGs under the assumption of Newtonian dynamics, $a_{\rm can}$. On the y-axis is the acceleration implied by the observed dynamics under the assumption of virial equilibrium, $a_{\rm dyn}$. The thin light red crosses show the data for the individual ETGs. The thick dark red line is for each galaxy the median in $a_{\rm dyn}$ of itself and its 20 closest neighbours in $a_{\rm can}$. An Exception are the ten ETGs with the lowest value in $a_{\rm can}$ and the ten ETGs with the highest value in $a_{\rm can}$, for which this quantity cannot be calculated, because this quantity should incorporate 10 ETGs with smaller radial accelerations and 10 ETGs with larger radial accelerations. The concept of the solid green lines is the same of the thick red line, except that they do not indicate the median, but the location of the forth most massive galaxy, and forth least massive galaxy, respectively, of their samples of 21 neighbouring ETGs in $a_{\rm dyn}$. The thin blue lines represent the different RARs discussed in this paper: the dotted blue line stands for Newtonian dynamics, the dashed blue line stands for Migromian dynamics with the simple $\mu$-function and the solid blue line stands for Migromian dynamics with the standard $\mu$-function. Right panel: The same as the left panel, but with $\log_{10}(a_{\rm dyn}/a_{\rm can})$ on the y-axis. This rescaling makes the offset between the data and the theoretical predictions better visible. In both panels, the mass of the stellar-mass BHs is based on equation~\ref{eq:blackhole} and the metallicity the ETGs have today.}
\end{figure*}

\begin{figure*}
\centering
\includegraphics[scale=0.88]{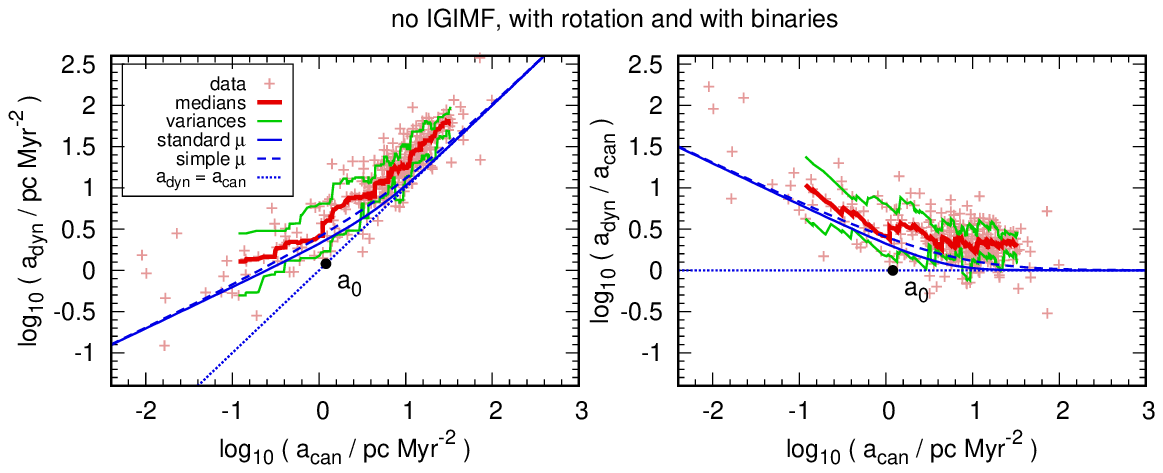}
\caption[The dynamical radial accelerations for ETGs over the expected acceleration based on $M_{\rm can}$ 2]{\label{fig:can-rot} As Fig.~\ref{fig:can-norot}, but with the dynamical accelerations of the ETGs adjusted for the presence of rotation and binaries in them.}
\end{figure*}

Fig.~\ref{fig:can-norot} shows the data on the dynamical radial accelerations of the 462 ETGs in our sample over their stellar radial accelerations. All ETGs are assumed her to have formed with the canonical IMF; i.e. without the changing IMFs according to the IGIMF-model. Also the effect of binaries, or that many of the ETGs also rotate, is not considered in the estimates of $a_{\rm dyn}$. The data are compared to the Newtonian RAR (equation~\ref{eq:RAR-newton}), the standard-MOND RAR (equation~\ref{eq:RAR-standard}), and the simple-MOND RAR (equation~\ref{eq:RAR-simple}). If one of these RARs encodes the dynamics of ETGs correctly, the data for the observed ETGs should be located along it. This is definitely not the case for the Newtonian RAR, which lies below most of the data. This is by orders of magnitude for low-mass ETGs with radial accelerations below the critical acceleration, $a_0$. Much of the data also lie above the Milgromian RARs, even though this is particularly at $a_{\rm can < a_0}$ much alleviated compared to the Newtonian RAR. Thus, unless non-baryonic dark matter is introduced, the data calls either for additional physical laws, for a decrease of the dynamical mass or an increase of the stellar mass including the stellar remnants of the ETGs. In old stellar systems like in the present sample, the latter can be the consequence of an overabundance of low-stars, an overabundance of high-mass stars, or both, compared to the canonical IMF.

The concept behind thin solid thin curves is the same as of the thick solid red line, except that the do not show the position of the ETGs with the eleventh-most $a_{\rm dyn}$ in their samples of 21 ETGs with neighbouring $a_{\rm dyn}$, i.e. their medians. Instead, they indicate the ETGs with the fourth-most $a_{\rm dyn}$, and the 18th-most $a_{\rm dyn}$, respectively. Alternatively, the lower green line is approximately the 16th percentile of ETGs with lower $a_{\rm dyn}$, the thick red line the 50th percentile, and the upper green line the 84th percentile. Thus, the two green lines indicate approximately the variance of the $a_{\rm dyn}$ of the sample of ETGs. However, this variance is not necessarily equal to the uncertainty of the measurements, but can in part be a real property of the sample itself. In other words, it is possible that not all of the variance would not go away, even if the errors to the measurements were infinitesimally small.

Fig.~\ref{fig:can-rot} shows a subset of 310 ETGs of the sample of 462 ETGs plotted in Fig.~\ref{fig:can-norot}, for which $M_{\rm dyn}^{\rm bin+rot}$ instead of $M_{\rm dyn}$ is shown on the y-axis. The data on the x-axis remains with $a_{\rm can}$ for every ETG the same as in Fig.~\ref{fig:can-norot}. However, the binary stars have almost no effect on $M_{\rm dyn}$ (see Section~\ref{sec:bin}), and the rotation of the ETGs still increases the mass needed to keep the ETGs in dynamical equilibrium. In other words, the disagreement of the data to the theoretical predictions from the RARs is even stronger than in Fig.~\ref{fig:can-norot}. On the other hand, rotation definitely takes place in many of the ETGs, even though its effect in Fig.~\ref{fig:can-rot} should perhaps be taken as an upper limit (see Section~\ref{sec:rot}). The most realistic values for the dynamical masses of the ETGs lie therefore probably in between the values shown in Fig.~\ref{fig:can-norot} (lower limit) and Fig.~\ref{fig:can-rot} (upper limit).

\subsection{ETGs with the IGIMF}
\label{sec:IGIMF-results}

\begin{figure*}
\centering
\includegraphics[scale=0.88]{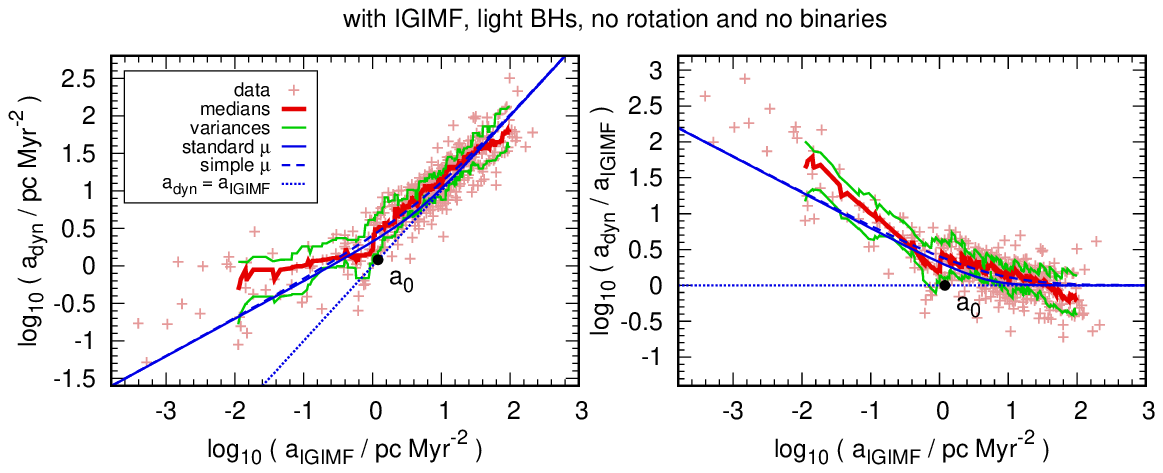}
\caption[The dynamical radial accelerations for ETGs over the expected acceleration based on $M_{\rm IGIMF}$ 3]{\label{fig:igimf-norot} As Fig.~\ref{fig:can-norot}, but with the stellar radial accelerations adjusted to the IGIMF (Sections~\ref{sec:IGIMF} to~\ref{sec:masses-ETGs}) instead of using the canonical IMF (equation~\ref{eq:IMF}) for star formation in every ETG. The masses of the stallar-mass BHs follow equation~\ref{eq:blackhole} for the metallicities that the ETGs have today. Note that binaries and rotation of the ETGs are {\it not} considered.}
\end{figure*}

\begin{figure*}
\centering
\includegraphics[scale=0.88]{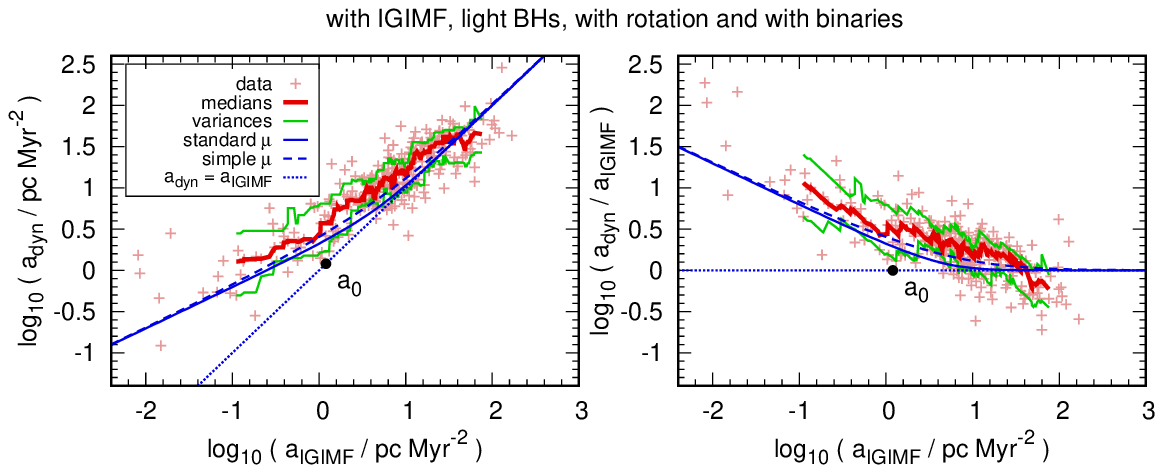}
\caption[The dynamical radial accelerations for ETGs over the expected acceleration based on $M_{\rm IGIMF}$ 4]{\label{fig:igimf-rot} As Fig.~\ref{fig:igimf-norot}, but with the dynamical accelerations of the ETGs adjusted for the presence of rotation and binaries in them.}
\end{figure*}

In contrast to Section~\ref{sec:canonical}, where the stellar radial accelerations of the ETGs were based on the canonical IMF (equation~\ref{eq:IMF}), they are in this this Section based on the simplified IGIMF (equation~\ref{eq:IGIMF3}; Section~\ref{sec:simplify}). 

As in Section~\ref{sec:dynmass}, the ETGs are separated in those with central velocity dispersions $\sigma_0 \le 100 \,$km/s and those with $\sigma_0 > 100 \,$km/s. 

For the ETGs with $\sigma_0 > 100 \,$km/s, the procedure of finding their IGIMF described in Sections~\ref{sec:IGIMF} to~\ref{sec:masses-ETGs} is applied only where equation~\ref{eq:MLpeak} produces values $\Upsilon >  \Upsilon_{\rm can}$, i.e. only for radii $R \apprle 0.388 R_{\rm e}$. The parameters in equation~\ref{eq:MLpeak} are for $\epsilon=1.29$ and $\eta=3.33$. For radii $R > 0.388 R_{\rm e}$, the IGIMF corresponds to the canonical IMF, indicating a moderate star formation like in the Milky Way there.

For ETGs with $\sigma_0 \le 100 \,$km/s, the recipe for finding their IGIMF is applied throughout the whole ETGs, without any radius restrictions.

Following the notion that the SFR follows matter density, the SFR should be changing within an ETG with radius, and consequently should its IGIMF. The ETGs with $\sigma_0 \le 100 \,$km/s are also those which have $M_{\rm IGIMF} \apprle 5 \cdot 10^9 \, {\rm M}_{\odot}$, and as such they have high-mass IGIMF-slopes $\alpha_{IGIMF} \apprge 2.3$ (see \citealt{Dabringhausen2019}). The mass difference between the two extreme cases, namely an ETG that has $\alpha_{\rm IGIMF} = 2.3$ up to $150 \, {\rm M}_{\odot}$ and the same ETG with a mass cutoff above $1 \, {\rm M}_{\odot}$ is about 20 percent. Consequently, the errors made by assuming a constant IGIMF instead of a varying one are comparatively small in this range.

However, for $M_{\rm IGIMF} > 5 \cdot 10^9  \, {\rm M}_{\odot}$, the mass gain with rising galaxy mass is increasingly fast compared to a canonical IMF. It might reach values of $\approx 2$ for the most massive ETGs, if the IGIMF in their outskirts was the same as in their centres (see figures~8 and~9 in \citealt{Dabringhausen2019}). Thus, allowing for a change in the IGIMF with radius is much more important for high-mass ETGs, and the overall increase in mass drops from $M_{\rm IGIMF}/M_{\rm can} \approx 2$ to $M_{\rm IGIMF}/M_{\rm can} \approx 1.4$.

In Fig.~\ref{fig:igimf-norot}, the masses of the stellar-mass black holes are thought to be influenced by the metallicities that the ETGs have today. They are given by equation~\ref{eq:blackhole}. The masses of the stellar-mass BHs become especially important in the high-mass ETGs, which had high SFRs, and therefore top-heavy IGIMFs. Thus, they have many stellar-mass BHs, compared to ETGs with the canonical IMF. The binaries and the rotation of the ETGs is neglected in Fig.~\ref{fig:igimf-norot}. The high-acceleration, and thus predominately high-mass ETGs fit much better to the theoretical Milgromian RARs in Fig.~\ref{fig:igimf-norot} than in Figs.~\ref{fig:can-norot} and~\ref{fig:can-rot}, where the star formation in the ETGs took place with the canonical IMF.

Fig.~\ref{fig:igimf-rot} considers also the binaries and the rotations of the ETGs (see Sections~\ref{sec:bin} and~\ref{sec:rot}), in contrast to Fig.~\ref{fig:igimf-norot}. All other parameters are the same as in Fig.~\ref{fig:igimf-norot}. Compared to Fig.~\ref{fig:igimf-norot}, considering rotation therefore increases the dynamical radial accelerations needed to keep the ETGs in equilibrium, and thus the dynamical masses. In the end, the values for $a_{\rm dyn}$ are too high on average here as well, except for the ETGs with the highest $a_{\rm dyn}$. Due the close correlation between masses and accelerations of the ETGs (see left panel of Fig.~\ref{fig:acc-mass}), this can be translated into a deficiency of $M_{\rm dyn}$ with growing $M_{\rm IGIMF}$ for ETGs with $M_{\rm IGIMF} \apprge 10^{11} \, {\rm M}_{\odot}$, to the point where $M_{\rm IGIMF} > M_{\rm dyn}$ on average. In contrast to $M_{\rm IGIMF} < M_{\rm dyn}$, which could be explained by undetected matter, $M_{\rm IGIMF} > M_{\rm dyn}$ indicates an overestimate of matter with our model in the most massive ETGs (see especially the end of this Section for an explanation).

\begin{figure*}
\centering
\includegraphics[scale=0.88]{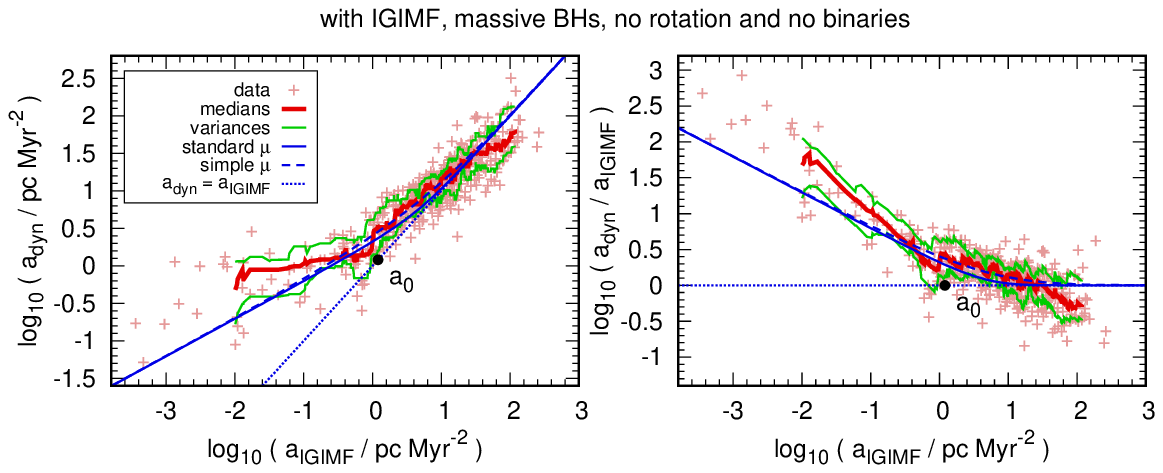}
\caption[The dynamical radial accelerations for ETGs over the expected acceleration based on $M_{\rm IGIMF}$ 5]{\label{fig:igimf-m-norot} As Fig.~\ref{fig:igimf-norot}, but with the masses of the stellar-mass BHs set to the initial mass of their progenitor stars, instead of that their masses follow equation~\ref{eq:blackhole} and the metallicities are as measured today. Note that binaries and rotation of the ETGs are {\it not} considered.}
\end{figure*}

\begin{figure*}
\centering
\includegraphics[scale=0.88]{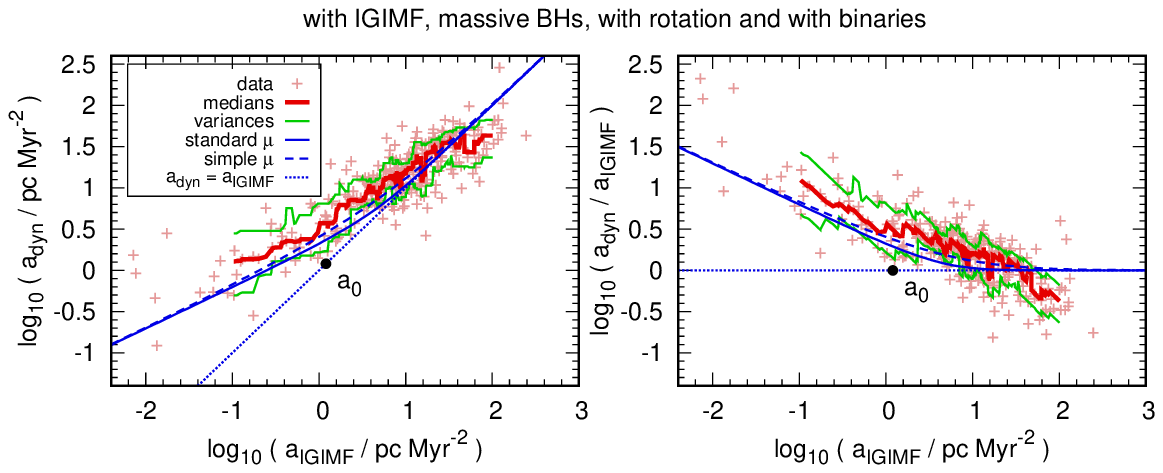}
\caption[The dynamical radial accelerations for ETGs over the expected acceleration based on $M_{\rm IGIMF}$ 6]{\label{fig:igimf-m-rot} As Fig.~\ref{fig:igimf-m-norot}, but with the dynamical accelerations of the ETGs adjusted for rotation and presence of binaries in them.}
\end{figure*}

\begin{figure*}
\centering
\includegraphics[scale=0.88]{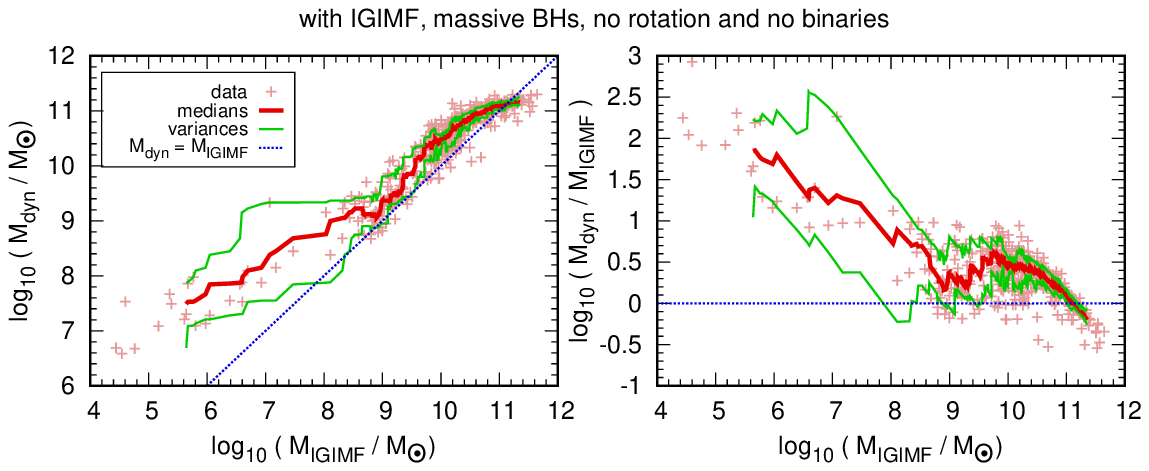}
\caption[The dynamical masses vs. their stellar masses of ETGs in the IGIMF-theory]{\label{fig:masses-dyn-IGIMF} The dynamical masses over their stellar masses for the ETGs in the IGIMF-model. Rotation and binary stars are not included here, and the masses of stellar-mass BHs equal the initial masses of their progenitor stars. The thick solid red curves are the medians of the ETGs. The thin solid green lines indicate the position of the ETGs with fourth-most $M_{\rm dyn}$, and 18th-most $M_{\rm dyn}$, respectively, in samples of 21 ETGs with neighbouring $M_{\rm dyn}$. Their concept is thus similar to the thick solid red curves, except that the red curves show the position in $M_{\rm dyn}$ of the ETGs with the eleventh-most $M_{\rm dyn}$. The dotted (blue) lines indicate equality between the two mass estimates.}
\end{figure*}

It is likely however that the metallicities the ETGs have today are higher than the metallicities under which most of the star formation in the massive ETGs took place. This is not modelled in this paper, but convincingly shown in Figure~2 in \citet{Jerabkova2018}. The masses of the stellar-mass BHs depend on the metallicities of the progenitor stars in such a way, that the more metal-poor the progenitor star is, the more massive the stellar-mass BH evolving out of it becomes; see e.g. figures~12 and~16 in \citet{Woosley2002}. These figures are however very imprecise, and only allow to name this general trend. What can be said though is that equation~\ref{eq:blackhole} most likely underestimates the real masses of stellar-mass BHs, even though it is only a very rough estimate by itself. Thus, we show in Figs.~\ref{fig:igimf-m-norot} and~\ref{fig:igimf-m-rot} also the results for the ETGs with the simplified IGIMF for the stellar-mass BHs having the initial mass of their progenitor stars. This is a strict upper limit, because real metal-poor massive stars also have stellar winds, albeit much less than their metal-rich counterparts.

Fig.~\ref{fig:igimf-m-norot} shows the data without considering binaries and rotation of the ETGs. Fig.~\ref{fig:igimf-m-rot} is the same as Fig.~\ref{fig:igimf-m-norot}, except that also the rotation and the binaries of the ETGs are considered. Thus, the data shown here are arguably more realistic than in Fig.~\ref{fig:igimf-m-norot}.

Fig.~\ref{fig:masses-dyn-IGIMF} shows the ETGs in terms of dynamical masses versus stellar masses instead of accelerations. The ETGs in this figure are plotted without considering rotation or binaries, and the masses of the stellar-mass BHs are roughly sketched based on the average metallicities that the ETGs have today. For the IGIMF-model with the assumptions made in this paper, the curve indicating the medians of the ETGs flattens in $M_{\rm dyn}$ for $M_{\rm IGIMF} \apprge 10^{10} \, {\rm M}_{\odot}$. For the ratios between $M_{\rm dyn}$ and $M_{\rm IGIMF}$, the curve goes downwards on average with increasing $M_{\rm IGIMF}$ until it nearly reaches $M_{\rm dyn}/M_{\rm IGIMF}=1$ at $M_{\rm IGIMF} \approx 10^9 \, {\rm M}_{\odot}$, then has upwards tendency for $10^9 \, {\rm M}_{\odot} \apprle M_{\rm IGIMF} \apprle 10^{10} \, {\rm M}_{\odot}$, and continues its downwards trend for $M_{\rm IGIMF} \apprge 10^{10} \, {\rm M}_{\odot}$. For $M_{\rm IGIMF} \apprge 10^{11} \, {\rm M}_{\odot}$,  $M_{\rm dyn} < M_{\rm IGIMF}$.

The last point indicates that the IGIMF-model, as it is applied here, very likely overestimates the mass in stars and stellar remnants for $M_{\rm IGIMF} \apprge 10^{11} \, {\rm M}_{\odot}$, because the ETGs cannot have $M_{\rm dyn} < M_{\rm IGIMF}$ systematically. It is not certain, however, whether such massive objects form in a single collapse, or rather as a merger of several smaller galaxies, as described e.g. in \citet{deLucia2007} for the most massive ETG in the centres of galaxy clusters in the $\Lambda$CDM-model, or \citet{Eappen2022} for the same in an Milgromian model. The dynamical masses of smaller galaxies do not undercut their stellar masses as severely, or may even be larger than the stellar masses, even in the simple model presented here. 

\subsection{Star formation times}
\label{sec:SFT}

\begin{figure}
\centering
\includegraphics[scale=0.88]{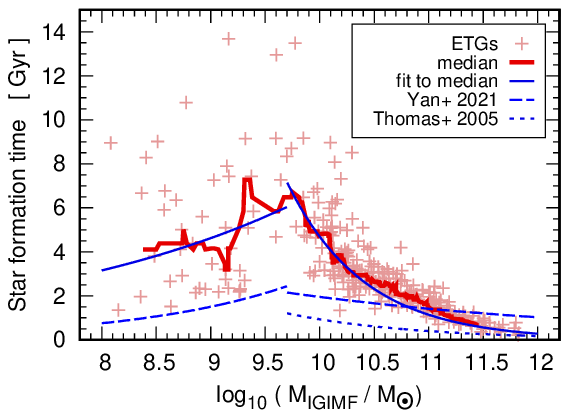}
\caption[The star formation times of the ETGs]{\label{fig:timescale} The star formation times of the ETGs with $a/a_0 > 0.1$. Shown is the time it takes for the ETGs to reach their $M_{\rm dyn}$ with the star formation rate as calculated from equation~\ref{eq:deltatMdyn} as thin light red crosses. The thick dark red line is for each galaxy the median in $M_{\rm dyn}$ of itself and its 20 closest neighbours in $M_{\rm can}$. An exception are the ten ETGs with the lowest value in $M_{\rm can}$ and the ten ETGs with the highest value in $M_{\rm can}$, for which this quantity cannot be calculated. This is because this quantity should incorporate 10 ETGs with even smaller $M_{\rm can}$, and 10 ETGs with even larger $M_{\rm can}$, respectively. The solid blue curve is the data for the ETGs fitted with equation~\ref{eq:parameters} for $M_{\rm IGIMF} \le 5\cdot 10^9 \, {\rm M}_{\odot}$ (i.e. the breaking point of the curve), and the same equation with a different set of parameters for $M_{\rm IGIMF} > 5\cdot 10^9 \, {\rm M}_{\odot}$. It has thus the same functional structure like the dashed blue curve, but different parameters. The dashed blue curve traces equation~\ref{eq:deltatMdyn}, and approximates the time is takes for the ETGs to reach their $M_{\rm IGIMF}$ with their SFR according to \citet{Yan2021}. Equation~\ref{eq:deltatMdyn} is equal to equation~3 in \citet{McDermid2015} for $M_{\rm IGIMF} > 5\cdot 10^9 \, {\rm M}_{\odot}$. Also equation~5 from \citet{Thomas2005} is shown for comparison as dotted blue curve.}
\end{figure}

The SFRs of the ETGs can be calculated based on the star formation times from \citet{Yan2021}, and the mass derived from the stellar content of the ETGs, $M_{\rm IGIMF}$. \citet{Yan2021} have calculated the star formation times for ETGs with $M_{\rm can} < 5 \cdot 10^9 \, {\rm M}_{\odot}$ themselves, but above this mass limit, the star formation times are the ones from \citet{McDermid2015}. The median values are shown as equations~18 and~19 in \citet{Yan2021}, or equation~\ref{eq:deltatMdyn} here. However, even though $M_{\rm IGIMF}$ is the mass created through star formation according to the IGIMF, is not necessarily equal to $M_{\rm dyn}$, which is the dynamical mass of the ETG. Especially for the more massive ETGs, which are not easily disturbed by tidal fields, $M_{\rm dyn}$ indicates the total amount of matter a ETG contains. If the ETGs formed with the IGIMF and with the calculated SFR, but without non-baryionic DM, then their star formation times should be as long as it takes for them to reach $M_{\rm dyn}$. These star formation times are shown in figure~\ref{fig:timescale} for ETGs with $a/a_0 > 0.1$. Also shown in this figure is a least-squares fit of these data with
\begin{equation}
	\label{eq:parameters}
	F(M_{\rm IGIMF})=\beta \, (M_{\rm dyn}/{\rm M}_{\odot})^{\gamma}.
\end{equation}
It is shown as a solid blue line in Fig.~\ref{fig:timescale}. For $M_{\rm IGIMF}/{\rm M}_{\odot} \le 5 \cdot 10^9 {\rm M_{\odot}}$, the parameters $\beta$ and $\gamma$ in Equation~\ref{eq:parameters} are $\beta = 0.148 \pm 0.135$ and $\gamma = 0.166 \pm 0.036$. For $M_{\rm IGIMF}/{\rm M}_{\odot} > 5 \cdot 10^9 {\rm M_{\odot}}$, they are $\beta = 5.50 \cdot 10^5 \pm 0.79 \cdot 10^5$ and $\gamma = -0.507 \pm 0.006$.

It is actually not important for the parameters found for Equation~\ref{eq:parameters} which one of the two extreme choices regarding the mass of the stellar-mass BHs introduced in Section~\ref{sec:masses-ETGs} are made. These are (1) an estimate of the BH-masses inspired by the metallicities that the ETGs have today and (2) assuming the initial mass of the progenitor stars for the BH mass. The parameters actually given in equation~\ref{eq:parameters} are for the arithmetic means of the two extremes. For all other parameter choices, the deviation is $\apprle 10^{-3}$ times the values given for Equation~\ref{eq:parameters}, or in other words, less than the line width of the solid curve in figure~\ref{fig:timescale}. 

Also equation~\ref{eq:deltatMdyn} is shown in this figure as a dashed blue line, which is where the SFRs, on which equation~\ref{eq:parameters} is based, originate. Thus, equation~\ref{eq:parameters} is not a completely independent estimate of the star formation times of the ETGs. Instead, if the star formation rate of an ETG would be given, then equation~\ref{eq:deltatMdyn} would give the time it would take the ETG to reach $M_{\rm IGIMF}$, and equation~\ref{eq:parameters} would give the time it would take the same ETG to reach $M_{\rm dyn}$. Equation~\ref{eq:deltatMdyn} is also the best fit to the data from {\sc GalIMF} in \citet{Yan2021}. Finally, equation~5 from \citet{Thomas2005} is shown for comparison.

For estimating the difference between $M_{\rm IGIMF}$ and $M_{\rm dyn}$, the simple $\mu$-function was taken, as it probably represents the data in the acceleration range of ETGs better than the standard $\mu$-function (e.g. \citealt{Milgrom2012,Chae2019slow}). ETGs with $a/a_0 < 0.1$ were omitted, because they probably suffer from tidal distortions to a large amount (see figures~\ref{fig:can-norot} to~\ref{fig:igimf-m-rot}).

In essence, equation~\ref{eq:parameters} is an estimate for the star formation times of the ETGs, based on their measured masses of the ETGs according to their dynamics. This approach confirms downsizing, i.e. less massive ETGs formed less rapidly. Incidentally, downsizing appears to be a consequence of the formation of ETGs through direct free-fall gravitational collapse of non-rotating gas clouds in the Milgromian framework \citep{Eappen2022}.

\section{Discussion}
\label{sec:discussion}

In summary, the Newtonian theoretical RAR does not fit to the observations of the ETGs in any of the Figs.~\ref{fig:can-norot} to~\ref{fig:igimf-m-rot}. Especially in the low-accelertion part, the deviance of the Newtonian RAR and the observations is extreme. Many authors have tried to solve this issue before by adding non-baryonic dark matter to the ETGs. For low-luminosity ETGs, corresponding mainly to low radial accelerations, this has been done for e.g. by \citet{Strigari2008,Wolf2010} and \citet{Ackermann2014}. For high-luminosity ETGs, corresponding mainly to high radial accelerations, this has been done for intance by \citet{Cappellari2006,Bolton2008} and \citet{Tortora2009}. A slightly better fit is achieved for the high-luminosity ETG with the Milgromian theoretical RARs. Both Newtonian dynamics and Milromian dynamics do however need the IGIMF-theory (or something equivalent) to fit the data in the high-luminosity range.

It is striking that $M_{\rm dyn} < M_{\rm IGIMF}$ on average for the highest values of $M_{\rm IGIMF}$. This could point to the simplicity of the model proposed here. For instance, it is assumed here that all ETGs formed an a similar fashion, independent of their mass. In truth however, the low-mass ETGs might form though monolithic collapses, while the most massive ETGs might form by mergers. The IGIMFs of the most massive ETGs would thus be more akin to their less massive counterparts than proposed here, because they were formed from such. Also, Fig.\ref{fig:timescale} shows the variety of parameters that different teams find in fitting an exponential function to the data of the star formation times of ETGs in dependence of their mass. However, even choosing an exponential function for this task was an ad-hoc choice in the first place. Moreover, the metallicity dependence was completely neglected in the model proposed here.

For extremely low radial accelerations, the ETGs tend to have higher dynamical radial accelerations than also the theoretical Milgromian RARs suggest. This discrepancy is however still lower of the order of one or two magnitudes than in Newtonian dynamics.

We will discuss these problems in the next Sections.

\subsection{Out-of-equilibrium dynamics}
\label{sec:no-virial} 

Out-of-equilibrium dynamics could well be the reason for extremely high mass-to-light ratios in ETGs \citep{Kroupa1997,Casas2012}.

However, the high-mass ETGs have mainly also large accelerations (left panel of Figure~\ref{fig:acc-mass}), and are therefore hard to disturb by external gravitational fields. This is not to say that it cannot be done, as examples of mergers of massive galaxies show (e.g. \citealt{Toomre1977}). Even if they were ETGs and in equilibrium before, they are irregular galaxies out of equilibrium during the interaction process. They are therefore unlikely to appear in the catalogue by \citet{Dabringhausen2016a}, or in this paper, even though they will eventually become ETGs. Thus, the high-mass ETGs in this paper can safely be assumed to be in virial equilibrium, and their elevated mass-to-light ratios must have other reasons (like the IGIMF discussed here).

The low-mass ETGs on the other hand, i.e. predominately the ETGs with luminosities $L_V \apprle 10^6 \ {\rm L}_{\odot}$, are very easy to disturb by external gravitational fields. This is because also their radial accelerations are low (left panel of Figure~\ref{fig:acc-mass}). The traces of these disturbances are tidal streams, which are however hard to detect because they are very faint, in contrast to the tidal streams of their more massive counterparts. Thus, low-mass ETGs are not necessarily excluded from the catalogue by \citet{Dabringhausen2016a}, because \citet{Dabringhausen2016a} decided on the appearance of the galaxy as an ETG (i.e. no clear tidal streams detected), and their classification as ETGs in the source literature. Also recall that low-mass ETGs, just as their more luminous counterparts, contain little gas \citep{Mateo1998}\footnote{There is one exception to this rule: The Leo~T dSph may contain more gas than stars \citep{RyanWeber2008}}. Thus, gas in general cannot be the reason for their high dynamical radial accelerations.

There is strong evidence that many, if not most low-mass ETGs are tidal dwarf galaxies (e.g. \citealt{Kroupa2010,Dabringhausen2013,Haslbauer2019}) and are as such basically free of non-baryonc dark matter (e.g. \citealt{Barnes1992a,Bournaud2010}). At least in the $\Lambda$CDM-model (and thus Newtonian dynamics), low-mass, dark-matter-free systems are vulnerable to tidal forces, as opposed to systems which are protected by the mass of their surrounding CDM-haloes \citep{Smith2015}. Tidal forces can thus increase the internal velocity dispersions of tidal dwarfs by additional pulls from the outside. Consequently, the estimates for the mass of the object is elevated if the (in this case wrong) assumption of tidal equilibrium is held \citep{Kroupa1997,Casas2012,Dominguez2016}. Even a system disupted by tides may still be detected as an over-density of stars, for which $\sigma_{\rm obs}$ and $R_{\rm e}$ can be observed. However, an estimate of the mass of the system based on equations~\ref{eq:Mdyn},~\ref{eq:Mdyn-bin},~\ref{eq:Mdyn-rot} or~\ref{eq:Mdyn-binrot} is clearly impossible. Also interactions with intergalactic gas may drive low-mass ETGs out of virial equilibrium \citep{Yang2014}.

Besides their (at least apparently) high mass-to-light ratios, there is also other evidence that low-mass ETGs are tidally disturbed. For instance, the ETGs around the Milky Way are generally more elliptical, the less massive and the closer to their host galaxy they are \citet{McGaugh2010}. More detailed observations of single galaxies show further peculiarities that can be interpreted as evidence for non-equilibrium dynamics for many of these ETGs, like elongated or irregular shapes, assymetic surface- brightness profiles, or asymmetric velocity-dispersion profiles (e.g. \citealt{Belokurov2006,Belokurov2007,Walker2009,Munoz2010,Willman2011,Deason2012,Asencio2022}). 

Thus, assuming Milgromian dynamics for low-mass ETGs implies that some of them are probably disturbed by tidal fields, but not necessarily all of them, as can be seen in Figs.~\ref{fig:can-norot} to~\ref{fig:igimf-m-rot}. This scenario appears more likely than the case that virtually {\it all} low-mass ETGs are out of virial equilibrium, as assuming Newtonian dynamics would suggest for them, unless they contain non-baryonic dark matter \citep{Dabringhausen2016b}.

Milromian dynamics was incorporated independently by \citet{Candlish2015} and by \citet{Lueghausen2015} into \textsc{RAMSES} by \citet{Teyssier2002}. It will be interesting to use these codes to compute the tidal processing of dwarf satellite galaxies in Milgromian dynamics, using the same approach as in \citet{Kroupa1997}.

\subsection{How symmetric is the data?}
\label{sec:symmetry}

How good is the assumption of complete isotropy of the {\it random} motions in the ETGs in Section~\ref{sec:dynmass}?

If the mass-to-light ratios are constant, the observed line-of-sight velocity distributions of over 2000 ETGs studied by \citet{Vudragovic2016} indicate that radial orbits are preferred over tangential orbits in them. Consequently, \citet{Vudragovic2016} find for their sample of ETGs that more realistic mass estimates based on the internal dynamics are on average about 10 per cent higher than expected under the assumption of isotropy. The noted discrepancy between mass estimates based on the internal dynamics of ETGs and mass estimates based on the amount of baryons detected in them is usually interpreted as a presence of non-baryonic dark matter.

On the other hand, \citet{Chae2019slow} have studied slowly rotating, massive ETGs from the ATLAS$^{\rm 3D}$-sample \citep{Cappellari2011}, which are also part of the sample presented here. They have shown that the need for anisotropic velocity dispersions is relaxed, if the mass-to-light ratio increases towards the centres of the ETGs. Moreover, \citet{Chae2019slow} use exactly the same parametrization of the gradient in the mass-to-light ratio as we do (Section~\ref{sec:dynmass}). Thus, in consequence, assuming that the {\it random} motion in ETGs is completely isotropic is perhaps not as far from the truth, as it would appear from \citet{Vudragovic2016}.

\section{Summary and conclusions}
\label{sec:conclusion}

If massive early-type galaxies (ETGs) formed with the canonical IMF (equation~\ref{eq:IMF}) and are in tidal equilibrium, they are generally too massive for their light by a factor of the order of two. There is in principle no reason why alternative stellar mass functions could not be the reason, except for the long-standing paradigm that the canonical IMF is also universal to all star formation \citep{Kroupa2001}. For the missing mass, non-baryonic dark matter is substituted (see e.g. \citealt{Cappellari2006,Bolton2008,Tortora2009}).

Here we propose an alternative by saying that ETGs have formed according to the IGIMF-theory (\citealt{Jerabkova2018} and references therein), and they obey Milgromian Dynamics, or MOdified Newtonian Dynamics (MOND) instead of Newtonian Dynamics. Finally, also the assumption of virial equilibrium might have to be relaxed in them. To do this, the ETGs are plotted with their observed dynamical accelerations versus their accelerations based on their presumed stellar populations. The results are then compared to different radial acceleration relations (RARs), i.e. predictions for the dynamical radial accelerations as functions of the baryons contained in the galaxies \citep{Wu2015,McGaugh2016,Lelli2017}.

This not to say that there were no attempts before to explain the properties of massive ETGs with IMFs different to the canonical one (equation~\ref{eq:IMF}). For instance, \citet{vanDokkum2010,Ferreras2013} and \citet{LaBarbera2013} had to use IMFs with more faint low-mass stars than the canonical IMF to explain the ratios of line indices in ETGs. The IMFs they proposed were however constant in time and space in each ETG. Later on, the independence of space in an ETG was relaxed by \citet{LaBarbera2016,vanDokkum2017,Parikh2018} and \citet{Sarzi2018}, who detected gradients in the IMFs of the ETGs they observed, such that the overabundance of low-mass stars was tied to the centres of the ETGs, while at their outskirts their IMFs approached the canonical IMF. \citet{Leier2016} confirmed from the line indices of the massive ETGs that they are over-abundant in low-mass stars compared to the canonical IMF, while the masses that these slopes suggest are still too light for the masses that gravitational lensing implies.

This puzzle may be resolved with an IMF that can change in space {\it and} time, like sketched in fig.~2 of \citet{Jerabkova2018} for the IGIMF-model. The IGIMF-model was introduced by \citet{Kroupa2003}, and further developed for the stellar populations of ETGs by \citet{Weidner2013a} and \citet{Weidner2013b} as a theory where the shape of the IMF depends on the star formation rate (SFR) and the metallicity of the star forming medium. An ETG, which is too massive for its contemporary IMF, can be understood in the IGIMF-model by having a high SFR in its past and therefore formed an overabundance of stars which have become NSs and BHs today, whereas today it is more metal-rich than the Milky Way and thus forms an over-abundance of low-mass stars. In consequence, we measure the light from the stellar population of today, but its total mass is boosted by burnt-out stars instead of non-baryonic dark matter. The advantage of the IGIMF is that it acts as an overarching theory to previous attempts to explain puzzling observations in ETGs with ad-hoc variations of the IMF with their mass, their SFR, et cetera. 

Whether Newtonian Dynamics or Milgromian dynamics is chosen is rather unimportant for high-luminosity ETGs, let alone which type of $\mu$-function describes the transition from the quasi-Newtonian regime to the fully Milgromian regime correctly. This is because the luminous ETGs also have generally high accelerations at their half-mass radius. In principle, this reflects the fact that any theory of gravity must result in Newtonian gravity in every-day situations, independent of its behaviour with extremely low accelerations, like in galaxies.

Also this is in principle well known. \citet{Tortora2014} already discussed the observed $\sigma_{\rm obs}$ and $M_{\rm obs}$ in the inner parts of 220 massive ETGs covered in the ATLAS$^{\rm 3D}$-survey \citep{Cappellari2011} under the premise of Milgromian dynamics instead of non-baryonic dark matter. They found that also in Milgromian dynamics, the observed $\sigma_{\rm obs}$ and $M_{\rm obs}$ are better explained with a Salpeter IMF, i.e. an IMF that has (in contrast to the canonical IMF) a slope of $\alpha_1=2.3$ instead of $\alpha_1=1.3$. \citet{Samurovic2014} confirms this trend in a very detailed study of 10 massive ETGs, which appear too heavy to have formed with the canonical IMF, independent of whether Newtonian or Milgromian dynamics was chosen.

It is noteworthy that the dynamical masses are lower than the stellar masses for the majority of the ETGs with the highest stellar masses (see Fig.~\ref{fig:masses-dyn-IGIMF}). This cannot be correct, because the dynamical mass of an ETG must encompass {\rm at least} the mass of its stellar population. Thus, either the assumed star formation rates are too high for the most massive ETGs in our simple model, or these ETGs are the products of mergers of several smaller ETGs, for which $M_{\rm dyn} > M_{\rm IGIMF}$. This might also be in our simplified model, which only treats the dependence of IGIMF on the SFR for high stellar masses, but not its dependence on metallicity for low stellar masses. However, the dependence on metallicity likely plays a minor role compared to the dependence on the SFR for the $M_{\rm dyn}/L_V$-ratios of the studied ETGs (see Section~\ref{sec:simplify}).

For low-acceleration ETGs in contrast, replacing Newtonian gravity with Milgromian gravity helps a great deal in explaining the apparent discrepancy between their expected accelerations based on their stellar populations mass and their observed dynamical mass based on their dynamics. However, also in Milgromian dynamics, the observed dynamical accelerations tend to be too high for the theorical radial acceleration relations (RARs). On the other hand, the missing-mass problem is much smaller than for the Newtonian RAR. It even dissapears in some low-acceleration ETGs, while not on averaging with windows that contain 21 ETGs. 

The slight over-estimate of the average dynamical radial accelerations of the low-acceleration ETGs in Milgromian dynamics can especially be seen in Figs.~\ref{fig:can-norot},~\ref{fig:igimf-norot} and~\ref{fig:igimf-m-norot}. These figures have the disadvantage that they neglect a possible rotation of the ETGs they show. The figures which do show an influence by rotation (Figs.~\ref{fig:can-rot},~\ref{fig:igimf-rot} and~\ref{fig:igimf-m-rot}) encompass only a subset of the ETGs shown in the full sample, and predominately the ETGs with low accelerations are missing because they are too faint, and thus challenging to observe. However, even if these low-acceleration ETGs would rotate, it would only increase the mass required to keep them bound, and the mismatch already seen in Figs.~\ref{fig:can-norot},~\ref{fig:igimf-norot} and~\ref{fig:igimf-m-norot} would become even larger.

This leads us to relax the assumption of virial equilibrium in the ETGs. For the high-luminosity ETGs, this is not an option, because they have also high radial accelerations and are therefore difficult to disturb from the outside. However, this may also be unnecessary for their dynamics, which could be explained by the IGIMF-theory. This is especially so, if the masses of the black holes they produced are rather high. This should not be a problem, because most stars, and thus also massive stars, probably formed in them when their metallicity was much lower than it is today (see figure~2 in \citealt{Jerabkova2018}). According to a comparison of figures~12 and~16 in \citet{Woosley2002}, low-metallicity stars have higher masses when they become stellar-mass black holes.

The low-luminosity ETGs on the other hand also have low radial accelerations, especially if they do not contain dark matter (be it baryonic or non-baryonic), but only their stellar populations. They are therefore much easier to disturb. The magnitude of this disturbance can also be explained in Newtonian dynamics, as is shown by \citet{Kroupa1997,Casas2012} and \citet{Dominguez2016}. They could well reach mass-to-light ratios of $10^4 {\rm M}_{\odot}/{\rm L}_{\odot}$ during their long-lived remnant phase, if they are (then wrongly) assumed to be in virial equilibrium. Note that dynamical mass-to-light ratios and dynamical radial accelerations (e.g. equation~\ref{eq:acc-obs}) are both quantities that are linear in dynamical mass. Thus, raising the dynamical mass by a factor of, say, 10 would incease the dynamical radial acceleration and the dynamical mass-to-light ratio by the same amount. It is highly doubtful however, if this would concern {\it all} low-luminosity ETGs, like Newtonian dynamics implies in Figs.~\ref{fig:can-norot} to~\ref{fig:igimf-m-rot}.

Milgromian dynamics on the other hand implies that some of them are probably disturbed by tidal fields, but not all of them, as can be seen in the same Figures. Also, the distorsions would have to be less extreme than in Newtonian dynamics. The scenario of Milgromian dynamics appears to be more likely, if ETGs are indeed free of non-baryonic dark matter. This would have to be the case, if low-mass ETGs are indeed tidal dwarf galaxies, like their arrangement in disks of satellites suggests \citep{LyndenBell1976,Kroupa2005,Ibata2013,Ibata2014,Mueller2016,Pawlowski2020}. This is because tidal dwarf galaxies do not contain non-baryonic dark matter, even if their progenitors did \citep{Barnes1992b,Bournaud2010}.

Thus, with all three ingredients together (i.e. the IGIMF, Migromiam dynamics and non-equilbrium dynamics), it may be possible to explain the dynamics of ETGs without dark matter. Without the IGIMF in particular, especially the dynamics of high-mass ETGs cannot be explained, as already noted in \citet{Dabringhausen2016b} with the same ETGs. This is remedied in this paper by allowing the $M_{\rm dyn}/L_V$-ratios of the massive ETGs to have a central peak, which is probably a consequence of the high star formation rates there in the past. This would have lead to a top-heavy IMF in the past and more remnants today according to the IGIMF-model. Also rotation of the ETGs is discussed, in contrast to \citet{Dabringhausen2016b}. On the other hand, it is {\it not} the purpose of the present paper to offer exact solutions to the missing mass of individual ETGs, but rather to show that it can be found with well estabished alternatives to non-baryonic dark matter. The IGIMF-theory may be a key ingredient to finding the missing mass. However, further research is needed to fully understand the dynamics of ETGs. Notable is that by using the IGIMF-theory and Milgromian gravitation (MOND) to explain the $M_{\rm dyn}/L_V$ values, we confirm that more massive ETGs form more rapidly (downsizing). This is in turn consistent with ETGs having formed through direct free-fall collapse of non-rotating gas clouds \citep{Eappen2022}.

\section*{Acknowlegdements}

J\"{o}rg Dabringhausen and Pavel Kroupa acknowledge support from the Grant Agency of the Czech Republic under grant number 20-21855S. They also thank the Deutscher Akademischer Austauschdienst-Eastern European exchange program at the University of Bonn for supporting the Bonn-Prague exchange. Finally they acknowlegde that particularly insightful suggestions by the anonymous referee improved the paper significantly. Throughout the preparation of this paper, they have made extensive use of NASA's Astrophysics Data System .

\section*{Data availability}

This paper utilizes the catalogue with the properties of early-type galaxies presented in \citet{Dabringhausen2016a}. No other data sources were used. 

\bibliographystyle{mn2e}
\bibliography{a-e,f-j,i-m,n-z}

\label{lastpage}

\section*{Appendix}

\begin{figure*}
	\centering
	\includegraphics[scale=0.88]{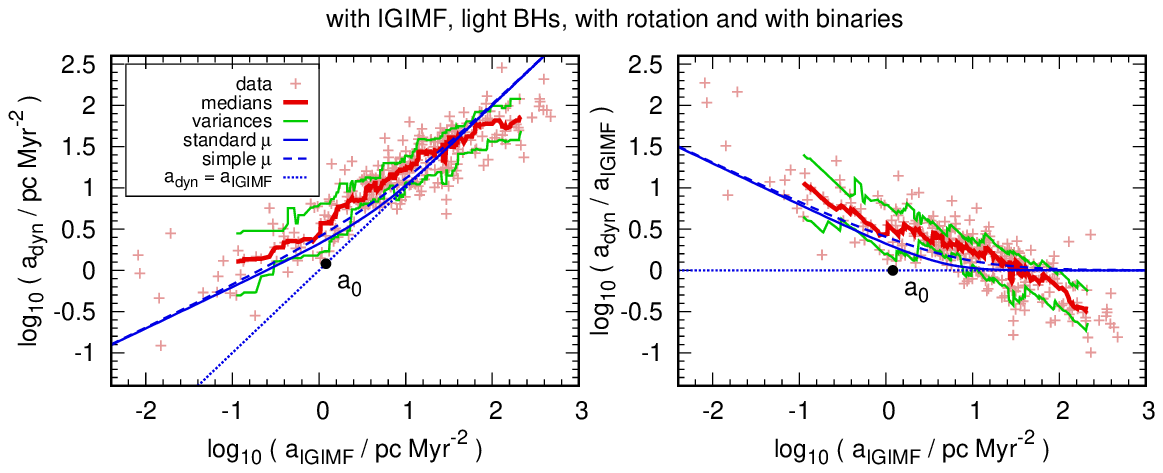}
	\caption[The dynamical radial accelerations for ETGs over the expected acceleration based on $M_{\rm IGIMF}$ 4]{\label{fig:igimf-rot-vD} As Fig.~\ref{fig:igimf-rot}, but with the higher increase in stellar mass resulting from the more extreme parameters $\epsilon=2.33$ and $\eta=6.0$ in equation \ref{eq:MLpeak} from \citet{vanDokkum2017}, rather than the standard case ($\epsilon=1.29$ and $\eta=3.33$) from \citet{Bernardi2018}.}
\end{figure*}

 For comparison with Figs.~\ref{fig:igimf-norot} to~\ref{fig:igimf-m-rot} in the main paper, Fig~\ref{fig:igimf-rot-vD} shows the more extreme choice made by \citet{vanDokkum2017} for the dependency of the mass-to-light ratio with the radius of the ETGs (i.e. $\epsilon=2.33$ and $\eta=6.0$ instead of $\epsilon=1.29$ and $\eta=3.33$ in equation~\ref{eq:MLpeak}). In this figure, the ETGs are assumed to rotate and to have black holes from metal-rich massive stars. Thus, the assumptions are identical to the ones in Fig.~\ref{fig:igimf-rot}, except for the relation describing the change of the mass-to-light ratio with radius. Consequently, even more ETGs have $M_{\rm dyn} < M_{\rm IGIMF}$ in this figure than in fig.~\ref{fig:igimf-rot}. Thus, the finding of \citet{Chae2018} and \citet{DominguezSanchez2019} that the standard case is more probable than the extreme case is corroborated here by using IGIMF-based stellar population synthesis.

\end{document}